\documentclass{emulateapj}
\usepackage{natbib}

\begin{document}

\title{Collisional debris as laboratories to study star formation}

\author{M. Boquien\altaffilmark{1,2,3}}
\email{boquien@astro.umass.edu}
\author{P.-A. Duc\altaffilmark{2}}
\author{Y. Wu\altaffilmark{4}}
\author{V. Charmandaris\altaffilmark{5,6}}
\author{U. Lisenfeld\altaffilmark{7}}
\author{J. Braine\altaffilmark{8}}
\author{E. Brinks\altaffilmark{9}}
\author{J. Iglesias-P\'aramo\altaffilmark{10}}
\author{C. K. Xu\altaffilmark{11}}

\altaffiltext{1}{University of Massachusetts, Department of Astronomy, LGRT-B 619E, Amherst, MA 01003, USA}
\altaffiltext{2}{AIM -- Unit\'e Mixte de Recherche CEA -- CNRS -- Universit\'e Paris VII -- UMR n$^\circ$ 7158}
\altaffiltext{3}{CEA-Saclay, DSM/DAPNIA/Service d'Astrophysique, 91191 Gif-sur-Yvette CEDEX, France}
\altaffiltext{4}{Astronomy Department, Cornell University, Ithaca, NY 14853, USA}
\altaffiltext{5}{Department of Physics, University of Crete, GR-71003, Heraklion, Greece}
\altaffiltext{6}{IESL/Foundation for Research and Technology - Hellas, GR-71110, Heraklion, Greece and Chercheur Associ\'e, Observatoire de Paris, F-75014, Paris, France}
\altaffiltext{7}{Dept. de F\'\i sica Te\'orica y del Cosmos, Universidad de Granada, Granada, Spain}
\altaffiltext{8}{Observatoire de Bordeaux, UMR 5804, CNRS/INSU, B.P. 89, F-33270 Floirac, France}
\altaffiltext{9}{Centre for Astrophysics Research, University of Hertfordshire, College Lane, Hatfield AL10 9AB, UK}
\altaffiltext{10}{Instituto de Astrof\'isica de Andaluc\'ia, Camino Bajo de Hu\'etor 50, 18008 Granada, Spain}
\altaffiltext{11}{California Institute of Technology, MC 405-47, 1200 East California Boulevard, Pasadena, CA 91125}

\begin{abstract}

In this paper we address the question whether star formation is driven by local processes or the large scale environment. To do so, we investigate star formation in collisional debris where the gravitational potential well and velocity gradients are shallower and compare our results with previous work on star formation in non-interacting spiral and dwarf galaxies. We have performed multiwavelength spectroscopic and imaging observations (from the far-ultraviolet to the mid-infrared) of 6 interacting systems, identifying a total of 60 star-forming regions in their collision debris.

Our analysis indicates that in these regions a) the emission of the dust is at the expected level for their luminosity and metallicity, b) the usual tracers of star formation rate display the typical trend and scatter found in classical star forming regions, and c) the extinction and metallicity are not the main parameters governing the scatter in the properties of intergalactic star forming regions; age effects and variations in the number of stellar populations, seem to play an important role. Our work suggests that local properties such as column density and dust content, rather than the large scale environment seem to drive star formation. This means that intergalactic star forming regions can be used as a reliable tool to study star formation.

\end{abstract}


\section{Introduction}

With the launch of the Spitzer and GALEX space observatories, the study of star formation in galaxies has seen tremendous development. Several surveys have been dedicated to star formation in the nearby universe. Among those, the Spitzer Infrared Nearby Galaxies Survey \cite[SINGS,][]{kennicutt2003a}, the Local Volume Legacy Survey \cite[LVL,][]{kennicutt2007b} and the Nearby Galaxy Survey \cite[NGS,][]{gildepaz2007a} have provided a large database of star-forming regions located in galaxies of various morphological types.

In galactic disks, star formation is governed by complex processes that may depend on the local and large-scale environments. The underlying old stellar population can affect star formation \citep{jog1984a,jog1984b} modifying the stability of molecular clouds. Gas clouds may collapse and form stars under the effect of density waves \citep{lin1964a}. In contrast, Galactic rotation provides a stabilizing effect \citep{toomre1964a}. In general, the interplay between the large scale and local processes is still a puzzle. In that respect, the dwarf irregular galaxies studied as part of the NGS and LVL surveys may be interesting laboratories: they lack density waves making them simpler objects than more massive galaxies. However, they are less chemically evolved; the low metallicity of their interstellar medium introduces a local difference in the star-formation process, with respect to spiral disks, which makes a direct comparison difficult. The outermost regions of isolated spiral galaxies have recently become popular alternative targets \citep{braine2004a,thilker2005a,gildepaz2005a}. Some systems show an UV excess, the origin of which has been strongly debated. These regions share the low metallicity of dwarf galaxies.

It is possible, however, to find at the same time a large scale environment similar to that of dwarfs and the local properties of spirals, i.e., the debris of galaxy-galaxy interactions. This debris is located outside massive galactic disks, but consist of chemically enriched material expelled from parent galaxies: the metallicity of their gas is close to solar \citep[see Fig.~17 in ][for instance]{duc2000a}\footnote{we use for the solar metallicity 12+log O/H=8.66 \citep{asplund2005a}.}. Furthermore, some metal-rich ``intergalactic'' regions may even be ``pure'', i.e. completely devoid of old stars, making them even less complex objects than typical dwarf galaxies.

Intergalactic star forming regions, in spite of having been proposed by \citet{zwicky1956a} and recognized as such by \citet{schweizer1978a}, have generated serious interest among astrophysicists for only about a decade. To put them better into context, intergalactic star forming regions can be classified as:

\begin{enumerate}
 \item Compact or diffuse emission-line regions, with no luminous stellar continuum \citep{weilbacher2003a,ryan2004a,mendes2004a,cortese2006a,werk2008a}. These faint star-forming regions are not gravitationally bound and contribute to the chemical enrichment of the intergalactic medium.
 \item Giant HII regions able to form young super star clusters, with masses between $10^6$~M$_\sun$ and $10^7$~M$_\sun$ \citep{weilbacher2003a,degrijs2003a,lopez2004a}.
 \item Tidal dwarf galaxies \citep[and references included in those papers]{duc1994a,duc2000a,braine2000a,boquien2007a,duc2007a}, i.e., gravitationally bound objects with sizes and masses similar to that of dwarf galaxies. They usually contain large quantities of gas in atomic and molecular form \citep{hibbard1994a,braine2000a, braine2001a,lisenfeld2002a,lisenfeld2004a}.
\end{enumerate}
Unlike star formation in disks, star formation in collisional debris may resemble ``beads on a string'' processes, that could be either stochastic \citep{smith2008a}, induced by gravitational instabilities \citep{elmegreen1996a} or a result of shocks \citep{struck1997a}.

To assess the potential of each of these processes, a preliminary study was performed by \citet{boquien2007a} on one system, NGC~5291, a very massive collisional ring hosting numerous star forming regions. We propose here to extend this work to a larger sample of objects. Six additional interacting systems exhibiting prominent intergalactic star forming regions were studied using multi-wavelength UV to mid-IR data.

In section 2 we present the selected systems, our PI and archival observations and data reduction. In section 3 we compare the properties of the star-forming regions in collisional debris to the more traditional process restricted to spiral galaxy disks. We discuss our results in section 4.

\section{Observations and data reduction\label{sec:observations}}

We have collected multi-wavelength data on a limited sample of carefully selected colliding systems, presented here below.

\subsection{Selected systems \label{ssec:systems}}

The goal of this paper is not to present a complete picture of intergalactic star-formation. The systems we have studied were mainly chosen based on the presence of prominent star forming regions in their collisional debris. As shown later, they are not representative of the Local Universe. Nevertheless, they offer the best laboratories where star formation (SF) can be studied in detail in this environment. All of them benefit from extensive multi-wavelength observations. In particular, they all have been observed in the UV by GALEX and in the mid-infrared by Spitzer. Besides the space-borne data, ground-based spectrophotometry is available. A high oxygen-abundance, typical of the outer regions of spiral disks (about half solar), has been measured in the HII regions throughout their collisional debris \citep{duc1994a,duc1998a,duc2000a,duc2007a}.

\begin{figure}
 \includegraphics[width=\columnwidth]{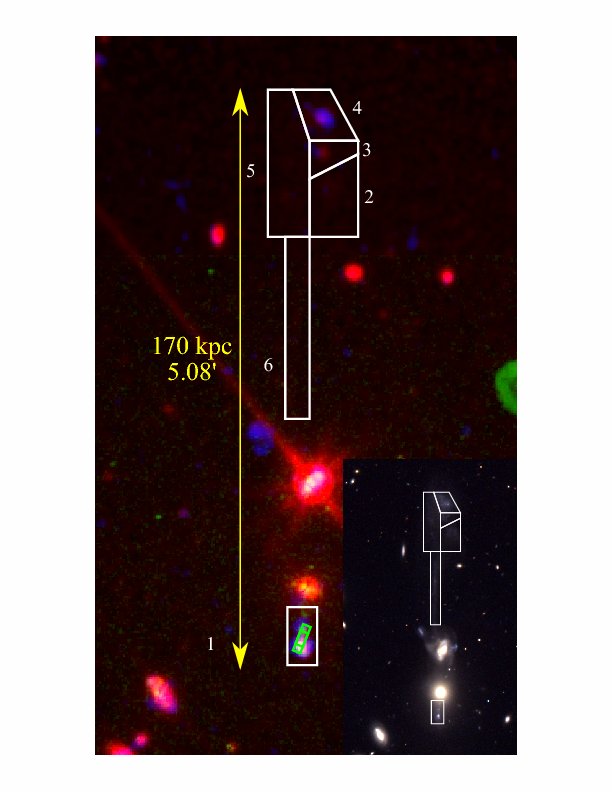}
 \caption{Arp~105 seen at 8.0~$\mu$m (red ; including the dust and the stellar emission), H$\alpha$ (green) and far ultraviolet (blue). North up, East left. The numbered white polygons are the apertures in which the fluxes have been measured. The green rectangles are the positions of the IRS slit. Inset image: optical BVR image taken at the CFHT 3.6m/12k.\label{fig:arp105}}
\end{figure}

Despite common characteristics --- active star-formation, high metallicity, environment ---, the six systems presented in this study turn out to present a variety of origins (early tidal interaction, on-going and final merger, direct collision) and environments (field, groups and cluster of galaxies). For most of them, numerical simulations able to account for their perturbed morphology have already been performed. Details on individual systems are given below in their description. Additional details are summarized in Table \ref{tab:pos-dist}. Optical images of all systems are shown in Fig. \ref{fig:arp105} to \ref{fig:vcc2062}.

\subsubsection{Arp~105}

Arp~105 (NGC~3561) is the result of an on-going collision between a spiral and an elliptical. This interaction has given birth to several tidal features: a 100~kpc long tidal arm (regions 2 to 6) to the north with knots of star formation at its tip (regions 3 and 4) apparently embedded in a low surface brightness older stellar structure, and a jet-like tail passing in front of the elliptical, ending (region 1) with a gas-rich, blue object \citep{duc1994a,duc1997a}. This young Tidal Dwarf Galaxy has the morphology of a Blue Compact Dwarf Galaxy (BCDG), while having an Interstellar Medium (Oxygen and CO-rich) more typical of a spiral galaxy.

\subsubsection{Arp~245}

Arp~245 (NGC~2992/93) is the result of an on-going collision between two spiral galaxies, connected by an apparent bridge \citep{duc2000a}. The tidal tail (regions 1 to 4) North of NGC~2992 ends (regions 1 and 2) with a star-forming complex where abundant molecular gas has been found \citep{braine2000a}. The color of the tail outside the star-forming regions (regions 3 and 4) indicates the presence of an old underlying stellar population.

\begin{figure}
 \includegraphics[width=\columnwidth]{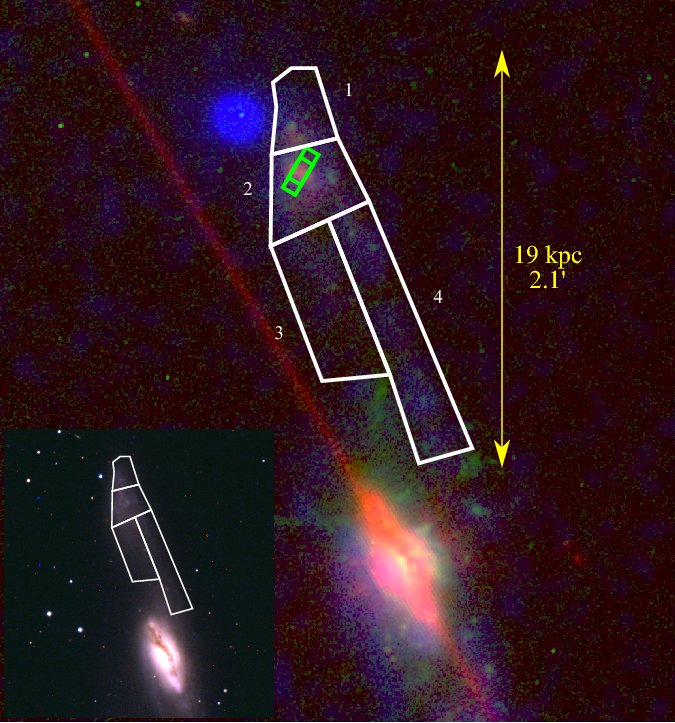}
 \caption{Arp~245 seen at 8.0~$\mu$m (red ; including the dust and the stellar emission), H$\alpha$ (green) and far ultraviolet (blue). North up, East left. The numbered white polygons are the apertures in which the fluxes have been measured. The green rectangles are the positions of the IRS slit. Inset image: optical BVR image taken at at ESO La Silla NTT/EMMI.\label{fig:arp245}}
\end{figure}

\subsubsection{NGC~5291}

NGC~5291 provides an example of a spectacular collisional ring around an early-type galaxy, probably resulting from the head-on, high-speed, collision with a massive companion within a cluster of galaxies \citep{bournaud2007a}. The giant (180kpc diameter), gas-rich, ring contains a string of metal-rich star-forming regions \citet{duc1998a}. The most luminous of them, in which CO lines \citet{braine2001a} and Polycyclic Aromatic Hydrocarbon (PAH) dust bands \citet{higdon2006a} were detected, are located within gravitationally bound, rotating, sub-structures with the mass of dwarf galaxies \citet{bournaud2007a}. Its intergalactic star-forming regions were studied in detail by \citet{boquien2007a}. In this paper we present new GALEX Far-Ultraviolet data. 

\begin{figure}
 \includegraphics[width=\columnwidth]{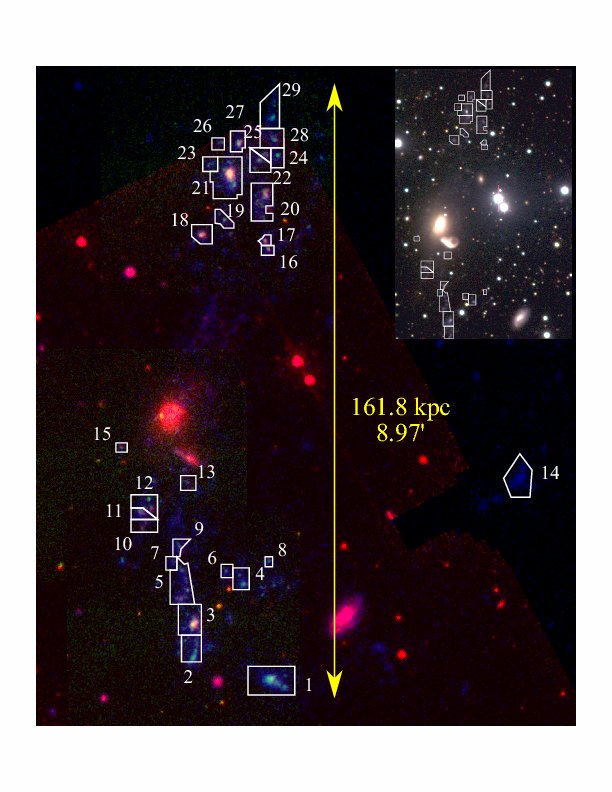}
 \caption{NGC~5291 seen at 8.0~$\mu$m (red ; including the dust and the stellar emission), H$\alpha$ (green) and far ultraviolet (blue). North up, East left. The numbered white polygons are the apertures in which the fluxes have been measured. Inset image: optical BVR image taken at at ESO La Silla NTT/EMMI. \label{fig:ngc5291}}
\end{figure}

A new far ultraviolet image is presented in Fig. \ref{fig:ngc5291}. See also Fig.~1 and 2 in \citet{boquien2007a}. The two brighest clumps, NGC~5291S and NGC~5291N correspond to regions 3 and 21.

\subsubsection{NGC~7252}

NGC~7252 is the archetypal example of an advanced merger of two spiral galaxies, observed between $5\times10^8$ and $10^9$ years after the initial collision \citep{hibbard1995a}. Each of its two tidal tails hosts one blue star-forming Tidal Dwarf Galaxy at its end. The one in the East contains two weak star forming regions (regions 3 and 4) whereas the North-West tail contains the brightest one (region 1).

\begin{figure}
 \includegraphics[width=\columnwidth]{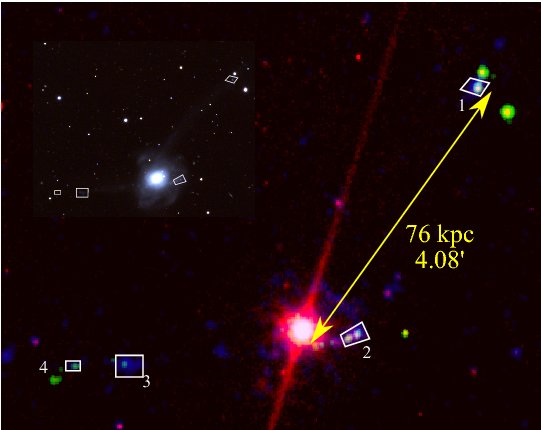}
 \caption{NGC~7252 seen at 8.0~$\mu$m (red ; including the dust and the stellar emission), H$\alpha$ (green) and far ultraviolet (blue). North up, East left. The numbered white polygons are the apertures in which the fluxes have been measured. Inset image: optical BVR image taken at ESO La Silla NTT/EMMI.\label{fig:ngc7252}}
\end{figure}

\subsubsection{Stephan's Quintet}

Stephan's Quintet is a well known compact group of galaxies that have suffered interactions of various types:

\begin{itemize}
\item tidal interactions, responsible for the formation of several tails, one particularly faint and old, another, more prominent, hosting a dust enshrouded, molecule-rich, star-forming object \citep[SQ-B;][]{lisenfeld2002a} (region 6) with another star forming region at its tip (region 2), and a recent one, made of pure atomic gas which has condensed locally to form faint but compact knots of star-formation, regions 3 to 5 \citep{mendes2004a}.
\item violent shocks due to an intruding galaxy which hit the system at a line of sight speed of 1000~km\,s$^{-1}$. These are thought to lie at the origin of H$\alpha$ and X-ray filaments where also prominent mid-infrared molecular H$_2$ lines have been detected \citep{appleton2006a}, regions 7 to 10 and 17). The shock region is most likely devoid of young stars and is excluded from this analysis. However, a luminous star-forming region known as SQ-A (regions 11 to 13), with PAH emission, has developed at its boundary.
\end{itemize}
Other intergalactic star forming regions are detected South-West of the system in regions 14 to 16.

\begin{figure}
 \includegraphics[width=\columnwidth]{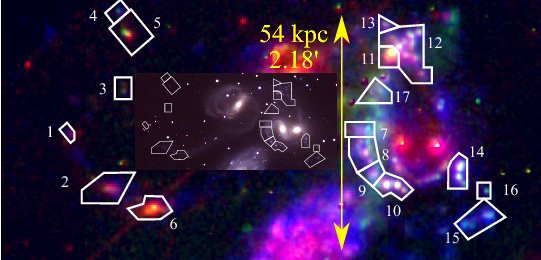}
 \caption{Stephan's Quintet seen at 8.0~$\mu$m (red ; including the dust and the stellar emission), H$\alpha$ (green) and far ultraviolet (blue). North up, East left. The numbered white polygons are the apertures in which the fluxes have been measured. Inset image: optical BVR image taken at Calar Alto 3.5m/MOSCA.\label{fig:SQ}}
\end{figure}

\subsubsection{VCC~2062}

Finally, VCC~2062 provides in the nearby Universe an example of a well resolved Tidal Dwarf Galaxy, which is likely the result of an old merger \citet{duc2007a}. Located in the Virgo cluster, it is linked to its putative parent galaxy, NGC~4694, by an HI arm which only shows an optical counterpart at the location of the dwarf galaxy. VCC~2062 exhibits several HII regions within a low-surface brightness stellar body. Its high oxygen abundance and strong CO detection is inconsistent with it being a classical dwarf galaxy.

\begin{figure}
 \includegraphics[width=\columnwidth]{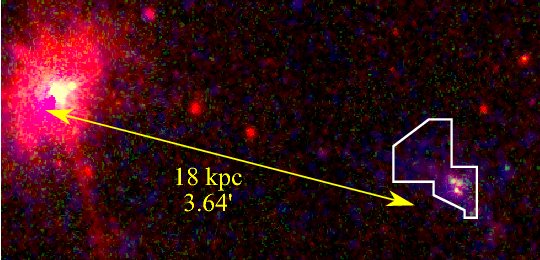}
 \caption{VCC~2062 seen at 8.0~$\mu$m (red ; including the dust and the stellar emission), H$\alpha$ (green) and far ultraviolet (blue). North up, East left. The white polygon is the aperture in which the fluxes have been measured.\label{fig:vcc2062}}
\end{figure}

\begin{table*}[!htbp]
\centering
\caption{Position and distance of the selected systems.\label{tab:pos-dist}}
\begin{tabular}{c c c c c c c c}
\hline\hline
System&RA&Dec&Distance&Spatial resolution for 1$\farcs$\\
&(J2000)&(J2000)&(Mpc)&(pc)\\\hline
Arp~105&$11^h11^m13.1^s$&$+28^\circ41'57\farcs$&115&557.5\\
Arp~245&$09^h45^m45.3^s$&$-14^\circ20'50\farcs$&31&150.3\\
NGC~5291&$13^h47^m24.5^s$&$-30^\circ24'25\farcs$&62&300.6\\
NGC~7252&$22^h20^m44.8^s$&$-24^\circ40'42\farcs$&64&310.3\\
Stephan~Quintet&$22^h35^m57.5^s$&$+33^\circ57'36\farcs$&85&412.1\\
VCC~2062&$12^h48^m00.0^s$&$+10^\circ58'15\farcs$&17&82.4\\
\hline
\end{tabular}
\tablenotetext{}{The coordinates have been obtained from the NASA/IPAC Extragalactic Database (NED).}
\end{table*}

\subsection{Reference sample}

The properties of the individual star-forming regions within collisional debris have been systematically compared to those located in the disks of four spirals. This broad-band reference sample consists of the following galaxies:

\begin{itemize}
 \item M51, a well studied spiral galaxy located at 8.2~Mpc \citep[see for instance][]{calzetti2005a}
 \item M81, the main galaxy of the M81 group, located at 3.63~Mpc. Its star forming regions have been studied by \citet{perez2006a}.
 \item Arp~24, an interacting system at 27.6~Mpc which hosts massive star clusters studied by \citet{cao2007a},
 \item Arp~82, an interacting system at 57~Mpc exhibiting a long tidal arm containing multiple star forming knots, as illustrated in \citet{hancock2007a}.
\end{itemize}
While Arp~24 and Arp~82 are interacting systems, they are still in an early phase. Contrary to all objects in the primary list, star formation still mainly occurs in the galaxy under the ``classical'' regime and not along a tidal tail. This explains why they are included in the reference sample.

Since the most luminous star-forming objects in collisional debris have the luminosities of dwarf galaxies, it is also relevant to make a comparison with the {\it integrated}, broad-band properties of individual galaxies. Large reference samples are available. In this study we made use of the following ones:

\begin{itemize}
 \item the 75 spiral galaxies of the SINGS survey \citep{kennicutt2003a,dale2007a}
 \item the 19 dwarf galaxies with $\textrm{M}_\textrm{B}>-19$ from the sample by \citet{rosenberg2006a},
 \item the 66 starburst galaxies from the sample by \citet{engelbracht2008a}.
\end{itemize}

As we also study the mid-infrared spectral properties of tidal debris we have gathered several samples having Spitzer IRS observations:

\begin{itemize}
 \item Galaxies from the BCDG sample of \citet{wu2006a},
 \item the starburst sample of \citet{engelbracht2008a},
 \item individual star forming regions in M101 \citep{gordon2008a}.
\end{itemize}

\subsection{Imaging}

Three star-forming tracers -- the UV, H$\alpha$ and mid-infrared emission -- were used for the present study. Their observation is described here and summarized in Table~\ref{tab:summary-obs}.

\begin{table*}[!htbp]
\centering
\footnotesize
\caption{Summary of available imaging observations for the selected systems.\label{tab:summary-obs}}
\begin{tabular}{c c c c c c}
\hline\hline
System&Type & Instrument & Band& Source\\
\hline
Arp~105&Imaging & GALEX & FUV& archive\\
"&" & " & NUV & "\\
"&Imaging & Spitzer IRAC & 3.6~$\mu$m& archive \\
"&" & " & 4.5~$\mu$m & "\\
"&" & " & 8.0~$\mu$m & "\\
"&Fabry-Perot Imaging & CFHT 3.6m/MOS/FP & H$\alpha$ & PI, \protect\cite{bournaud2004a} \\\hline
Arp~245&Imaging & GALEX & FUV & archive\\
"&" & " & NUV & "\\
"&Imaging & Spitzer IRAC & 3.6~$\mu$m & archive\\
"&" & " & 4.5~$\mu$m & "\\
"&" & " & 8.0~$\mu$m & "\\
"&Narrow-band Imaging & La Silla NTT/EMMI & H$\alpha$ & PI, \protect\citet{bournaud2004a} \\\hline
NGC~5291&Imaging & GALEX & FUV & GI PID 32, \protect\citet{boquien2007a}\\
"&" & " & NUV & "\\
"&Imaging & Spitzer IRAC & 3.6~$\mu$m & archive, \protect\citet{boquien2007a}\\
"&" & " & 4.5~$\mu$m & "\\
"&" & " & 8.0~$\mu$m & "\\
"&Fabry-Perot Imaging & La Silla 3.6m/CIGALE & H$\alpha$ & PI, \protect\citet{bournaud2004a} \\\hline
NGC~7252&Imaging & GALEX & FUV & archive\\
"&" & " & NUV & "\\
"&Imaging & Spitzer IRAC & 3.6~$\mu$m& archive\\
"&" & " & 4.5~$\mu$m & "\\
"&" & " & 8.0~$\mu$m & "\\
"&Narrow-band imaging & KPNO & H$\alpha$ & archive \\\hline
SQ&Imaging & GALEX & FUV & archive, \protect\citet{xu2005a}\\
"&" & " & NUV & "\\
"&Imaging & Spitzer IRAC & 3.6~$\mu$m & archive\\
"&" & " & 4.5~$\mu$m & "\\
"&" & " & 8.0~$\mu$m & "\\
"&Narrow-band imaging & Calar Alto 2.1m & H$\alpha$ & PI, \protect\citet{xu1999a}\\\hline
VCC~2062&Imaging & GALEX & FUV & PI, \protect\citet{duc2007a}\\
"&- & " & NUV & "\\
"&Imaging & Spitzer IRAC & 3.6~$\mu$m& archive\\
"&" & " & 4.5~$\mu$m & "\\
"&" & " & 8.0~$\mu$m & "\\
"&Narrow-band imaging & KPNO & H$\alpha$ & archive, \protect\citet{duc2007a}\\\hline
\end{tabular}
\end{table*}

\subsubsection{Ultraviolet imaging}

All ultraviolet images were obtained using the GALEX space telescope. It offers 2 bands, FUV ($\lambda_\textrm{eff}$=151~nm) and NUV ($\lambda_\textrm{eff}$=227~nm) and a large field of view of 1.24$^\circ$. NGC~5291 observations were obtained through our GALEX program (PID 32). A NUV image of the system was presented in \citet{boquien2007a}. The FUV image which was not available at the time is now included in this paper. Maps of other systems where obtained from the NGS survey. Note that the Arp~105 image was taken by the GALEX All-sky Imaging survey and therefore the exposure time was limited to 115~s, instead of at least 500~s, as for the other systems. Due to the low photons count in ultraviolet, the sky histogram closely follows a Poisson distribution and the quantization is seen very clearly. To determine the sensitivity we have fitted the sky background histogram with a Poisson distribution of the form : $f\left(k;\lambda\right)=\lambda^k\exp\left(-\lambda\right)/k!$, where $k$ is the number of photons and $\lambda$ the variance. In the case the photon count is high enough to have a normal distribution, we have used the \textsc{msky} procedure\footnote{\textsc{msky} is an \textsc{IRAF} routine written by M. Dickinson (1993, private communication). It allows to define the interval in the pixel distribution where the mode and the variance are calculated.} to obtain the sensitivity per pixel. We have converted to a sensitivity per arcsec$^{2}$ mutiplying by the pixel scale (1.5\arcsec per pixel). The exposure times and the sensitivities for each band are summarized in Table \ref{tab:summary-fuv}.

\begin{table*}[!htbp]
\centering
\footnotesize
\caption{Summary of the ultraviolet observations of the selected systems.\label{tab:summary-fuv}}
\begin{tabular}{c c c c}
\hline\hline
System& Band & Exposure time & 1 $\sigma$~sensitivity \\
&&(s)&($\mu$Jy arcsec$^{-2}$)\\
\hline
Arp~105& FUV & 115 & $0.10$\\
"&NUV & " & $0.10$\\
Arp~245& FUV & 1052 &$0.05$\\
"& NUV & 4298 & $0.03$\\
NGC~5291& FUV & 5567 &$0.03$\\
"& NUV & 2886 & $0.02$\\
NGC~7252& FUV & 562 & $0.05$\\
" & NUV & " & $0.05$\\
SQ& FUV & 3322& $0.05$\\
"& NUV & " & $0.04$\\
VCC~2062 & FUV & 1530 & $0.03$\\
" & NUV & " & $0.03$\\\hline
\end{tabular}
\end{table*}

\subsubsection{H$\alpha$ imaging}

H$\alpha$ images were obtained using Fabry-Perot (FP) and narrow-band observations obtained at various ground based telescopes (see Table~\ref{tab:summary-obs}). The reconstructed FP images were flux calibrated using the narrow-band images as well as slit spectroscopy. See \citet{boquien2007a} for details. 

The exposure times and the sensitivities for each band are summarized in Table \ref{tab:summary-ha}. The sensitivities have been calculated using \textsc{msky}. The sensitivity could not be measured for NGC~7252. Indeed for this system, only archival H$\alpha$ images for which the background has been removed are available.

\begin{table*}[!htbp]
\centering
\footnotesize
\caption{Summary of the H$\alpha$ observations of the selected systems.\label{tab:summary-ha}}
\begin{tabular}{c c c}
\hline\hline
System& Exposure time& 1 $\sigma$~sensitivity\\
&(s)&($\mu$Jy arcsec$^{-2}$)\\
\hline
Arp~105& 300 &$3.3\times10^{-21}$\\
Arp~245& 900 &$5.7\times10^{-21}$\ \\
NGC~5291& 180--255 & $2.2\times10^{-21}$\\
NGC~7252&1200 & -- \\
SQ& 600 & $3.4\times10^{-21}$\\
VCC~2062& 1200 & $3.0\times10^{-21}$\\\hline
\end{tabular}
\end{table*}

\subsubsection{Mid-infrared imaging}

All infrared observations were taken from the Spitzer space telescope archives. For our systems, only images at 3.6, 4.5, 5.8 and 8.0~$\mu$m from the Infrared Array Camera (IRAC) were available. The MIPS 24~$\mu$m images of NGC~7252 and VCC~2062 are too shallow to be of use in this study.

The images were retrieved from the archives and processed (cosmic ray rejection, interpolation, mosaicking, etc.) using the standard IRAC pipeline. Due to the presence of strong gradients in the 5.8~$\mu$m images that would have made flux measurements particularly uncertain, they have not been used in this study. The sensitivities were calculated using \textsc{msky}. The exposure times and the sensitivities for each band are summarized in Table \ref{tab:summary-ir}.

\begin{table*}[!htbp]
\centering
\footnotesize
\caption{Summary of the mid-infrared observations of the selected systems.\label{tab:summary-ir}}
\begin{tabular}{c c c c}
\hline\hline
System& Band & Exposure time & 1 $\sigma$~sensitivity\\
&&(s)&($\mu$Jy arcsec$^{-2}$)\\
\hline
Arp~105& 3.6~$\mu$m& 450 & $0.12$ \\
"& 4.5~$\mu$m & " & $0.17$\\
"& 8.0~$\mu$m & " & $1.29$\\
Arp~245& 3.6~$\mu$m& 300 & $0.48$ \\
"& 4.5~$\mu$m & " & $0.56$\\
"& 8.0~$\mu$m & " & $1.96$\\
NGC~5291& 3.6~$\mu$m& 432 &$0.27$ \\a
"& 4.5~$\mu$m & " & $0.30$\\
"& 8.0~$\mu$m & " & $1.10$\\
NGC~7252& 3.6~$\mu$m& 120 & $0.23$ \\
"& 4.5~$\mu$m & " & $0.31$\\
"& 8.0~$\mu$m & " & $1.31$ \\
SQ& 3.6~$\mu$m& 432 & $0.20$ \\
"& 4.5~$\mu$m & " & $1.11$\\
"& 8.0~$\mu$m & " & $1.11$\\
VCC~2062& 3.6~$\mu$m& 120 & $0.20$ \\
"& 4.5~$\mu$m & " & $0.25$\\
"& 8.0~$\mu$m & " & $1.49$\\\hline
\end{tabular}
\end{table*}

The stellar emission from the 8.0~$\mu$m fluxes has not been subtracted. The main reason why any stellar subtraction was not attempted was that such a correction applied to intergalactic regions would increase the errors rather than reducing them. Indeed, these regions were chosen to have specifically a high SFR and a low background old stellar population. The stellar continuum subtraction is usually done using the IRAC 3.6~$\mu$m band. Assuming a [3.6]-[8.0]=0 color for the stars, as in \citet{pahre2004a}, the stellar contribution to the 8.0~$\mu$m flux is expected to be at about 7\% for our objects. However, in luminous SF regions the IRAC 3.6~$\mu$m band may be significantly polluted by the 3.3~$\mu$m PAH line. This contribution is very uncertain as the line wavelength is too short to be observed with IRS. Taking this into account, we have decided not to correct the 8.0 microns flux for the stellar contribution.

\subsubsection{Aperture photometry}

Aperture photometry was carried out on a number of selected star-forming regions in each system. All images had previously been registered to a common grid. The standard selection of the star forming regions is based on the presence of ultraviolet, H$\alpha$ and ultraviolet emission.

The apertures are defined to enclose as few star forming regions as possible in order to limit the number of stellar populations. The aim is to keep the star forming regions simple. Indeed the presence of an increasing number of stellar populations renders the study more complex. The limiting factor to define the apertures is the resolution of the infrared and ultraviolet images. To take into account the variety of resolutions we define the apertures taking into account all images simultaneously.

The background level is defined by averaging typically 5 to 20 measurements of the background around each star forming region. In case it is not possible to measure the background for each region individually\footnote{This happens generally when the star forming regions are too close to each other.}, the same background level is defined for a set of regions. The number of regions is kept as small as possible in order to limit the influence of any large scale background variability.

The procedure how the regions were selected and apertures defined is described in detail in \citet{boquien2007a}. The selected apertures are displayed in Fig.~\ref{fig:arp105} to \ref{fig:vcc2062}.

The measured fluxes are listed in Tables~\ref{tab:flux-arp105} to \ref{tab:flux-vcc2062}.

\begin{table*}[!htbp]
\centering
\caption{Ultraviolet, infrared and H$\alpha$ fluxes for the selected regions of Arp~105.\label{tab:flux-arp105}}
\begin{tabular}{c c c c c c c}
\hline\hline
Region&F$_\textrm{FUV}$&F$_\textrm{NUV}$&F$_\mathrm{3.6 \mu m}$&F$_\mathrm{4.5 \mu m}$&F$_\mathrm{8.0 \mu m}$&F$_{\mathrm{H}\alpha}$\\
&($\mu$Jy)&($\mu$Jy)&($\mu$Jy)&($\mu$Jy)&($\mu$Jy)&($10^{-19}$ W~m$^{-2}$)\\
\hline
$1$&$58\pm3$&$105\pm7\phantom{0}$&$489\pm192$&$348\pm129$&$2564\pm712$&$276\pm2$\\
$2$&$\phantom{0}1\pm5$&$\phantom{00}9\pm12$&$440\pm46\phantom{0}$&$270\pm57\phantom{0}$&$\phantom{0~}334\pm151$&--\\
$3$&$\phantom{0}5\pm3$&$\phantom{00}9\pm5\phantom{0}$&$200\pm17\phantom{0}$&$130\pm21\phantom{0}$&$\phantom{0~}373\pm56\phantom{0}$&--\\
$4$&$28\pm4$&$\phantom{0}72\pm8\phantom{0}$&$620\pm30\phantom{0}$&$390\pm37\phantom{0}$&$\phantom{0~}871\pm97\phantom{0}$&--\\
$5$&$\phantom{0}0\pm6$&$\phantom{0}12\pm16$&$730\pm59$&$470\pm73\phantom{0}$&$\phantom{0~}662\pm195$&--\\
$6$&$\phantom{0}8\pm8$&$\phantom{0}41\pm20$&$860\pm76\phantom{0}$&$490\pm94\phantom{0}$&$\phantom{0~}442\pm251$&--\\
\hline
\end{tabular}
\tablenotetext{}{The fluxes have not been corrected for Galactic extinction.}
\tablenotetext{}{The Spitzer IRS observation of Arp~105S is best associated with region 1.}
\end{table*}

\begin{table*}[!htbp]
\centering
\caption{Ultraviolet, infrared and H$\alpha$ fluxes for the selected regions of Arp~245.\label{tab:flux-arp245}}
\begin{tabular}{c c c c c c c}
\hline\hline
Region&F$_\textrm{FUV}$&F$_\textrm{NUV}$&F$_\mathrm{3.6 \mu m}$&F$_\mathrm{4.5 \mu m}$&F$_\mathrm{8.0 \mu m}$&F$_{\textrm{H}\alpha}$\\
&($\mu$Jy)&($\mu$Jy)&($\mu$Jy)&($\mu$Jy)&($\mu$Jy)&($10^{-19}$ W~m$^{-2}$)\\
\hline
$1$&$24\pm2$&$46\pm1$&$1350\pm31$&$\phantom{0}810\pm59\phantom{0}$&$2540\pm194$&$\phantom{0}73\pm8\phantom{0}$\\
$2$&$54\pm3$&$93\pm2$&$3630\pm51$&$2200\pm99\phantom{0}$&$9140\pm325$&$212\pm13$\\
$3$&$15\pm4$&$42\pm3$&$3360\pm74$&$2120\pm144$&$2980\pm476$&$\phantom{0}96\pm19$\\
$4$&$41\pm5$&$69\pm3$&$3510\pm93$&$2270\pm182$&$4530\pm601$&$145\pm24$\\
\hline
\end{tabular}
\tablenotetext{}{The fluxes have not been corrected for Galactic extinction.}
\tablenotetext{}{The Spitzer IRS observation of Arp~245N is best associated with region 2.}
\end{table*}

\begin{table*}[!htbp]
\centering
\caption{Ultraviolet, infrared and H$\alpha$ fluxes for the selected regions of NGC~5291.\label{tab:flux-ngc5291}}
\begin{tabular}{c c c c c c c}
\hline\hline
Region&F$_\textrm{FUV}$&F$_\textrm{NUV}$&F$_\mathrm{3.6 \mu m}$&F$_\mathrm{4.5 \mu m}$&F$_\mathrm{8.0 \mu m}$&F$_{\mathrm{H}\alpha}$\\
&($\mu$Jy)&($\mu$Jy)&($\mu$Jy)&($\mu$Jy)&($\mu$Jy)&($10^{-19}$~W~m$^{-2}$)\\
\hline
$1$&$255\pm18$&$274\pm20$&$331\pm102$&$254\pm35$&$640\pm99$&$332\pm101$\\
$2$&$82\pm6$&$92\pm7$&$144\pm41$&$98\pm24$&$629\pm41$&$107\pm33$\\
$3$&$134\pm10$&$160\pm12$&$385\pm24$&$251\pm32$&$2088\pm56$&$188\pm57$\\
$4$&$44\pm4$&$53\pm4$&$131\pm13$&$118\pm16$&$207\pm28$&$43\pm13$\\
$5$&$86\pm7$&$98\pm8$&$221\pm28$&$179\pm36$&$805\pm64$&$59\pm19$\\
$6$&$13\pm1$&$14\pm2$&$31\pm6$&$28\pm7$&$97\pm13$&$15\pm5$\\
$7$&$11\pm1$&$14\pm2$&$53\pm12$&$44\pm5$&$73\pm12$&$6\pm3$\\
$8$&$6\pm1$&$6\pm1$&$31\pm7$&$31\pm3$&$60\pm7$&$10\pm4$\\
$9$&$20\pm2$&$23\pm2$&$84\pm15$&$61\pm7$&$176\pm16$&$15\pm5$\\
$10$&$11\pm1$&$14\pm2$&$112\pm27$&$95\pm10$&$278\pm27$&$14\pm5$\\
$11$&$30\pm3$&$34\pm3$&$199\pm18$&$138\pm8$&$384\pm19$&$25\pm9$\\
$12$&$28\pm3$&$31\pm3$&$229\pm34$&$169\pm13$&$509\pm34$&$43\pm13$\\
$13$&$10\pm1$&$12\pm2$&$188\pm19$&$135\pm8$&$311\pm19$&$13\pm5$\\
$14$&$57\pm5$&$62\pm5$&--&--&--&--\\
$15$&$4\pm1$&$5\pm1$&$70\pm9$&$50\pm5$&$189\pm10$&$8\pm4$\\
$16$&$12\pm1$&$14\pm2$&$50\pm11$&$41\pm6$&$286\pm12$&$14\pm5$\\
$17$&$13\pm1$&$12\pm1$&$52\pm9$&$37\pm6$&$377\pm11$&$13\pm5$\\
$18$&$27\pm2$&$30\pm3$&$146\pm29$&$104\pm13$&$1047\pm30$&$33\pm10$\\
$19$&$16\pm2$&$18\pm2$&$96\pm19$&$51\pm8$&$267\pm19$&$16\pm5$\\
$20$&$73\pm6$&$87\pm7$&$287\pm59$&$244\pm22$&$1428\pm59$&$78\pm25$\\
$21$&$218\pm16$&$259\pm19$&$638\pm39$&$534\pm52$&$5236\pm90$&$463\pm139$\\
$22$&$25\pm2$&$30\pm3$&$63\pm27$&$34\pm16$&$427\pm28$&$21\pm7$\\
$23$&$14\pm2$&$17\pm2$&$14\pm18$&--&$114\pm17$&$15\pm5$\\
$24$&$21\pm2$&$21\pm2$&$2\pm21$&$25\pm11$&$31\pm20$&$19\pm7$\\
$25$&$12\pm1$&$15\pm2$&$38\pm13$&$30\pm8$&$235\pm14$&$12\pm5$\\
$26$&$7\pm1$&$9\pm1$&--&--&--&$7\pm3$\\
$27$&$36\pm3$&$44\pm4$&$115\pm23$&$100\pm14$&$614\pm24$&$42\pm13$\\
$28$&$30\pm3$&$33\pm3$&--&--&--&$23\pm9$\\
$29$&$38\pm3$&$41\pm4$&--&--&--&$25\pm9$\\
\hline
\end{tabular}
\tablenotetext{}{The fluxes have not been corrected for Galactic extinction.}
\tablenotetext{}{The Spitzer IRS observation of NGC~5291S (resp. NGC~5291N) is best associated with region 3 (resp. 21).}
\end{table*}

\begin{table*}[!htbp]
\centering
\caption{Ultraviolet, infrared and H$\alpha$ fluxes for the selected regions of NGC~7252.\label{tab:flux-ngc7252}}
\begin{tabular}{c c c c c c c}
\hline\hline
Region&F$_\textrm{FUV}$&F$_\textrm{NUV}$&F$_\mathrm{3.6 \mu m}$&F$_\mathrm{4.5 \mu m}$&F$_\mathrm{8.0 \mu m}$&F$_{\mathrm{H}\alpha}$\\
&($\mu$Jy)&($\mu$Jy)&($\mu$Jy)&($\mu$Jy)&($\mu$Jy)&($10^{-19}$~W~m$^{-2}$)\\
\hline
$1$&$32\pm2$&$32\pm2\phantom{0}$&$138\pm31\phantom{0}$&$87\pm24$&$\phantom{0}946\pm50\phantom{0}$&$93\pm47$\\
$2$&$36\pm4$&$40\pm13$&$221\pm256$&$194\pm126$&$1082\pm636$&$55\pm28$\\
$3$&$23\pm2$&$28\pm3\phantom{0}$&$113\pm26\phantom{0}$&$83\pm22$&$\phantom{0}231\pm92\phantom{0}$&$5\pm3$\\
$4$&$\phantom{0}4\pm1$&$\phantom{0}5\pm1\phantom{0}$&$\phantom{0}14\pm7\phantom{00}$&$\phantom{0}16\pm7\phantom{00}$&$\phantom{00}41\pm26\phantom{0}$&$9\pm5$\\
\hline
\end{tabular}
\tablenotetext{}{The fluxes have not been corrected for Galactic extinction.}
\end{table*}

\begin{table*}[!htbp]
\centering
\caption{Ultraviolet, infrared and H$\alpha$ fluxes for the selected regions of Stephan's Quintet.\label{tab:flux-SQ}}
\begin{tabular}{c c c c c c c c c c}
\hline\hline
Region&F$_\textrm{FUV}$&F$_\textrm{NUV}$&F$_\mathrm{3.6 \mu m}$&F$_\mathrm{4.5 \mu m}$&F$_\mathrm{8.0 \mu m}$&F$_{\textrm{H}\alpha}$\\
&$\mu$Jy&$\mu$Jy&$\mu$Jy&$\mu$Jy&$\mu$Jy&$10^{-19}$~W~m$^{-2}$\\
\hline
$1$&$2\pm1$&$2\pm1$&$31\pm10$&$15\pm5$&$225\pm17$&--\\
$2$&$16\pm2$&$19\pm2$&$215\pm58$&$127\pm32$&$1192\pm110$&$45\pm35$\\
$3$&$2\pm1$&$1\pm1$&$18\pm4$&$6\pm11$&$167\pm35$&$23\pm11$\\
$4$&$4\pm1$&$4\pm1$&$3\pm3$&$0\pm9$&$99\pm29$&$14\pm9$\\
$5$&$3\pm1$&$4\pm1$&$40\pm13$&$33\pm4$&$780\pm53$&$92\pm31$\\
$6$&$8\pm1$&$11\pm1$&$396\pm5$&$279\pm19$&$3541\pm65$&$149\pm22$\\
$7$*&$40\pm8$&$51\pm8$&$457\pm123$&$215\pm134$&$896\pm254$&$431\pm14$\\
$8$*&$61\pm15$&$82\pm14$&$928\pm225$&$459\pm245$&$1613\pm465$&$578\pm25$\\
$9$*&$33\pm7$&$43\pm7$&$326\pm103$&$161\pm112$&$689\pm212$&$308\pm12$\\
$10$*&$48\pm11$&$60\pm11$&$639\pm171$&$329\pm185$&$2100\pm352$&$576\pm19$\\
$11$&$19\pm4$&$23\pm6$&$550\pm159$&$360\pm86$&$3736\pm228$&$540\pm12$\\
$12$&$124\pm15$&$139\pm24$&$987\pm732$&$476\pm397$&$5246\pm1051$&$959\pm55$\\
$13$&$10\pm2$&$10\pm3$&$10\pm76$&$21\pm33$&$283\pm109$&$56\pm6$\\
$14$&$72\pm8$&$74\pm4$&$444\pm457$&$192\pm277$&$1654\pm287$&$399\pm18$\\
$15$&$60\pm5$&$60\pm5$&$144\pm188$&$77\pm150$&$391\pm203$&$177\pm30$\\
$16$&$5\pm2$&$5\pm2$&$89\pm41$&$50\pm33$&$82\pm45$&$36\pm7$\\
$17$*&$26\pm3$&$28\pm5$&$74\pm146$&$56\pm140$&$776\pm164$&$488\pm17$\\
\hline
\end{tabular}
\tablenotetext{}{The fluxes have not been corrected for Galactic extinction.}
\tablenotetext{}{The entries marked with an asterisk indicate shocked regions.}
\end{table*}

\begin{table*}[!htbp]
\centering
\caption{Ultraviolet, infrared and H$\alpha$ fluxes for the selected regions of VCC~2062.\label{tab:flux-vcc2062}}
\begin{tabular}{c c c c c c c}
\hline\hline
Region&F$_\textrm{FUV}$&F$_\textrm{NUV}$&F$_\mathrm{3.6 \mu m}$&F$_\mathrm{4.5 \mu m}$&F$_\mathrm{8.0 \mu m}$&F$_{\textrm{H}\alpha}$\\
&$\mu$Jy&$\mu$Jy&$\mu$Jy&$\mu$Jy&$\mu$Jy&$10^{-19}$~W~m$^{-2}$\\
\hline
1&$35\pm3$&$46\pm4$&$210\pm52$&$185\pm55$&$1597\pm387$&$42\pm15$\\
\hline
\end{tabular}
\tablenotetext{}{The fluxes have not been corrected for Galactic extinction.}
\end{table*}

\subsection{Mid-infrared spectroscopy\label{sssec:ir-spec}}

\begin{figure}
 \includegraphics[width=\columnwidth]{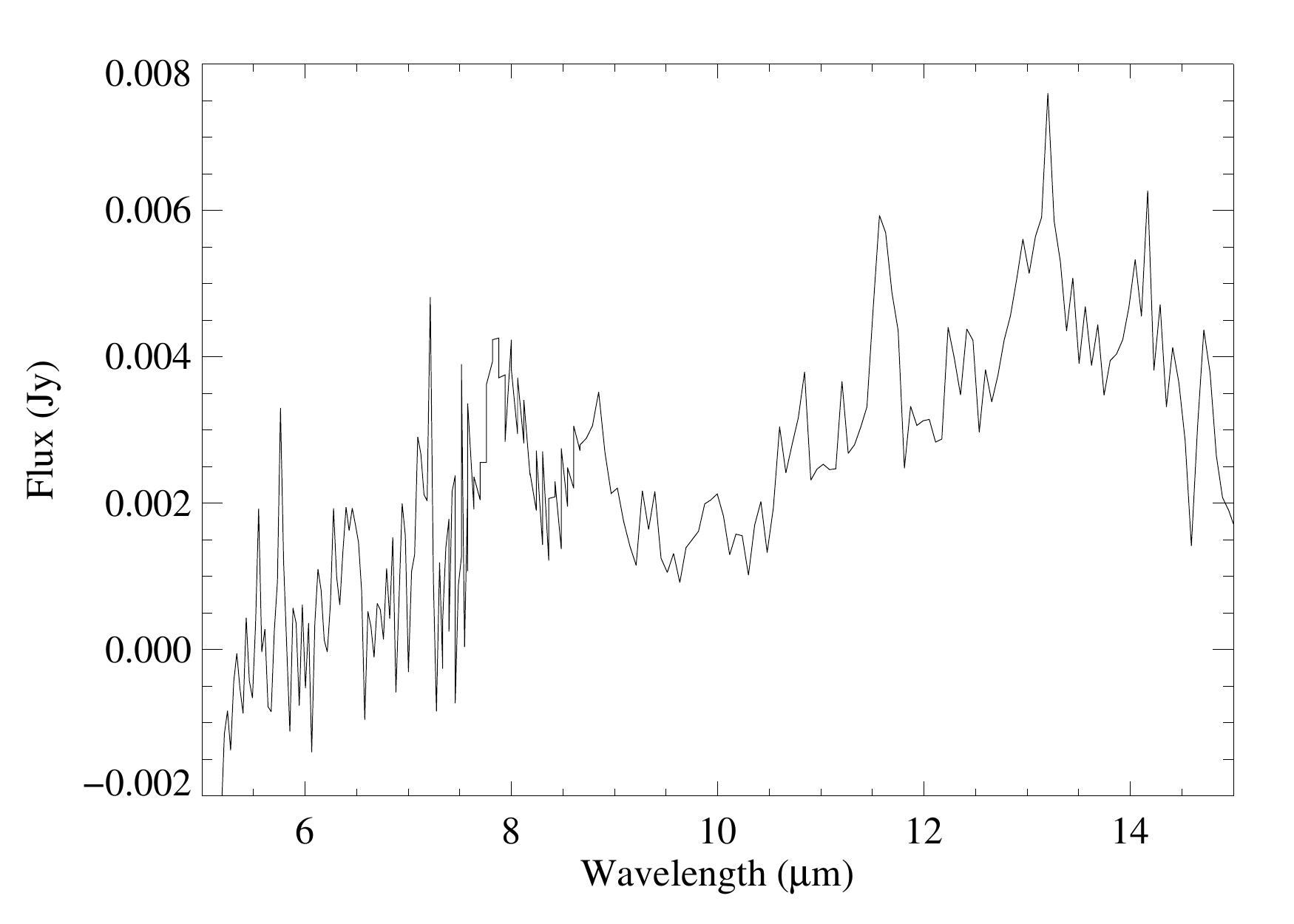}
 \caption{5-15 $\mu$m low-resolution spectrum of ARP~105S obtained with Spitzer/IRS. The strong PAH bands at 6.1, 7.7, 8.6 and 11.2 $\mu$m are visible. No rest-frame correction has been performed.\label{fig:arp105-IRS}}
\end{figure}

\begin{figure}
 \includegraphics[width=\columnwidth]{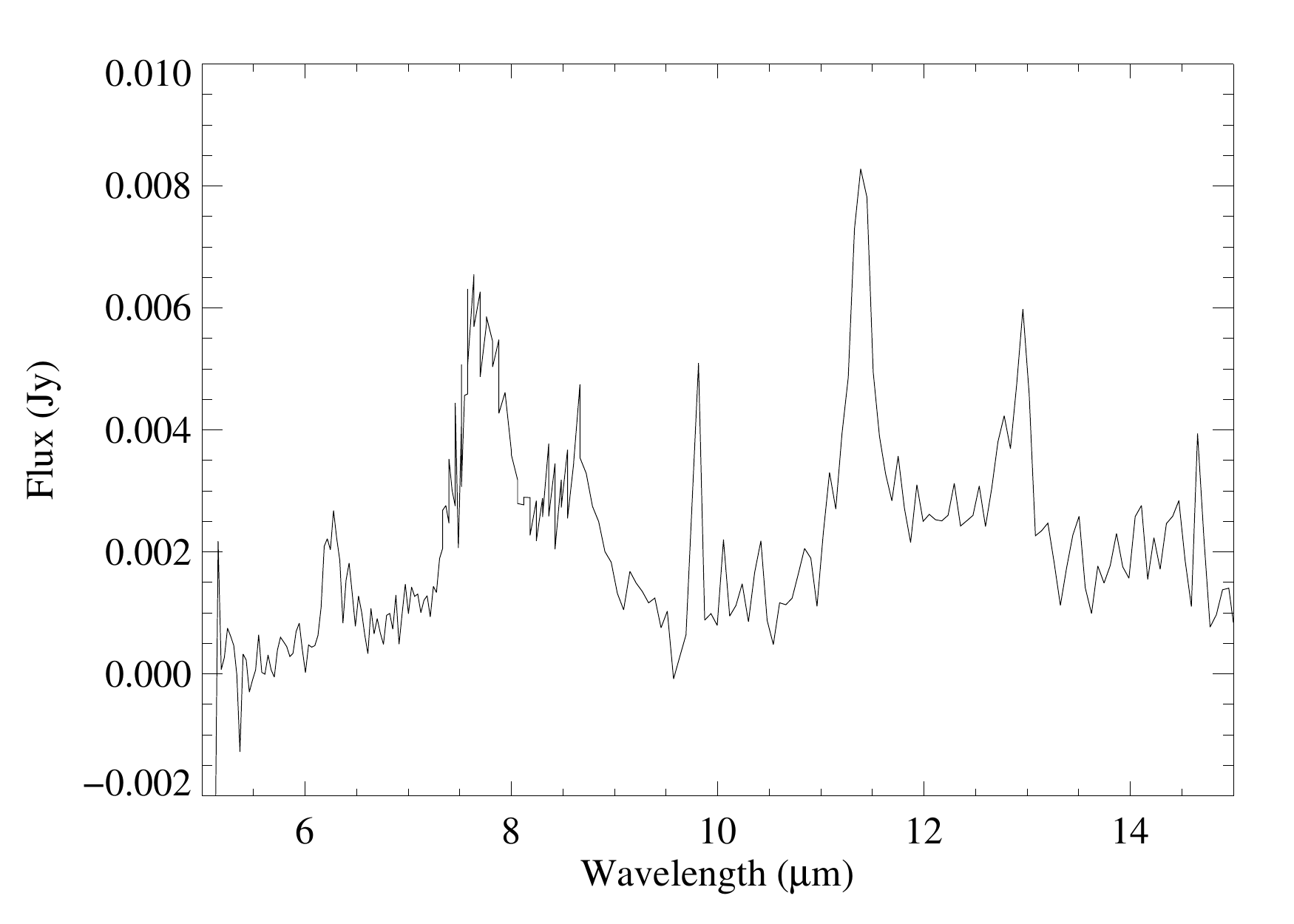}
 \caption{5-15 $\mu$m low-resolution spectrum of ARP~245S obtained with Spitzer/IRS. The strong PAH bands at 6.1, 7.7, 8.6 and 11.2 $\mu$m are visible. No rest-frame correction has been performed.\label{fig:arp245-IRS}}
\end{figure}

Unpublished archival Spitzer/IRS mid-infrared spectroscopic observations are presented here of two TDG candidates in Arp~105 and Arp~245. These observations were carried out using the Short-Low (SL: 5.2-14.5\,$\mu$m; R=64$\sim$128) and Short-High (SH: 9.9-19.6\,$\mu$m; R=600) modules of the Infrared Spectrograph IRS \citep{Houck2004a}. Arp~105S and Arp~245N were observed for 3 cycles of 60 seconds in the SL mode and 3 cycles of 120 seconds in the SH mode on December 12, 2005 and on May 30, 2005, respectively. The data were processed with the Spitzer Science Center (SSC) pipeline version S13.0.1. The reduction started from the intermediate pipeline products (droop files), which only lacked stray light and flat-field correction. Individual points to each nod position of the slit were co-added using median averaging. Sky background from the SL spectral data was taken out by differencing the images of the two orders in this module. Then the differenced images were extracted with the spectral modeling, analysis, and reduction tool \citep[SMART,][]{higdon2004a} using a variable width aperture, which scales the extraction aperture with wavelength to recover the same fraction of the diffraction-limited instrumental point-spread-function. The SH data were extracted with the full slit extraction method in SMART. Finally, the spectra were flux-calibrated by multiplying by the relative spectral response function (RSRF), which was created from the IRS standard stars $\alpha$Lac for SL and $\xi$Dra for SH for which accurate templates are available \citep{Cohen2003a}. The SL spectra are presented in Fig. \ref{fig:arp105-IRS} for Arp~105S (corresponding to region 1 in Fig. \ref{fig:arp105}) and \ref{fig:arp245-IRS} for Arp~245N (corresponding to region 2 in Fig. \ref{fig:arp245})

In Tables \ref{tab:IRS-EW-PAH} and \ref{tab:IRS-ionic}, we present the measurements of broad and narrow emission line features from the IRS spectra of Arp~105S and Arp~245N. We have also measured the equivalent width (EW) of NGC~5291N and NGC~5291S independently of \citet{higdon2006a}. Indeed, this allows an easier comparison with the data of \citet{wu2006a} whose EW have been measured using the same method. The EW of the PAH features are measured from the SL spectra. The PAH EW were derived by integrating the flux of the feature in the mean spectra of both nod positions above an adopted continuum and then dividing by the average adopted continuum. The baseline was determined by fitting a spline function to selected points. The wavelength limits for the integration of the features were approximately 5.95-6.55\,$\mu$m for the 6.2\,$\mu$m PAH, 7.15-8.20\,$\mu$m for the 7.7\,$\mu$m PAH, 8.20-8.90\,$\mu$m for the 8.6\,$\mu$m PAH and 10.80-11.80\,$\mu$m for the 11.2\,$\mu$m PAH features. The integrated fluxes for the fine structure lines of [SIV]\,10.51$\mu$m, [NeII]\,12.81$\mu$m, [NeIII]\,15.55$\mu$m and [SIII]\,18.71$\mu$m were measured by fitting a Gaussian profile using the line-fitting function in SMART. Note that no sky background has been subtracted from the SH spectra. This has not affected our measurements of the spectral lines, however.

\begin{table*}[!htbp]
\centering
\caption{Measured EW of the most prominent PAH bands.\label{tab:IRS-EW-PAH}}
\begin{tabular}{c c c c}
\hline\hline
Region&EW$_{6.2 \mu m}$&EW$_{7.7 \mu m}$&EW$_{11.2 \mu m}$\\
&($\mu$m)&($\mu$m)&($\mu$m)\\
\hline
Arp~105S&$0.865\pm0.473$&$0.756\pm0.249$&$0.245\pm0.109$\\
Arp~245N&$1.312\pm0.621$&$0.726\pm0.175$&$0.884\pm0.285$\\
NGC~5291N&$0.569\pm0.068$&$0.349\pm0.017$&$0.574\pm0.053$\\
NGC~5291S&$0.550\pm0.244$&$0.383\pm0.079$&$0.780\pm0.225$\\
\hline
\end{tabular}
\end{table*}

\begin{table*}[!htbp]
\centering
\caption{IRS fluxes of mid-infrared ionic lines\label{tab:IRS-ionic}}
\begin{tabular}{c c c c c c c}
\hline\hline
Region&F$_{\left[\textrm{S IV}\right]@10.51\mu m}$&F$_{\left[\textrm{Ne II}\right]@12.81\mu m}$&F$_{\left[\textrm{Ne III}\right]@15.55\mu m}$&F$_{\left[\textrm{S III}\right]@18.71\mu m}$\\
&($10^{-18}$~W~m$^{-2}$)&($10^{-18}$~W~m$^{-2}$)&($10^{-18}$~W~m$^{-2}$)&($10^{-18}$~W~m$^{-2}$)\\
\hline
Arp~105S&$5.41\pm1.16$&$4.71\pm0.58$&$9.47\pm0.44$&$9.49\pm2.47$\\
Arp~245N&--&$4.19\pm1.21$&$5.24\pm0.48$&$4.99\pm0.61$\\
\hline
\end{tabular}
\end{table*}

\section{Results\label{sec:results}}

As shown in Fig.~\ref{fig:arp105} to \ref{fig:vcc2062}, the UV, H$\alpha$ and 8$\mu$m maps of our selected interacting systems match well spatially, tracing the presence of a global process: star formation. However, within a given system, large fluctuations may be found in the relative fluxes from one individual region to the other, indicating local differences. Since by far the most complex and ambiguous tracer of star-formation is the mid-infrared emission, we detail in the following the dust properties of the collisional debris.

\subsection{Dust in collisional debris}

Mid-infrared PAH emission at 8.0~$\mu$m is observed within all the collisional debris of our sample; it is enhanced in or close to HII regions, as traced by the H$\alpha$ maps.

\begin{figure}
 \includegraphics[width=\columnwidth]{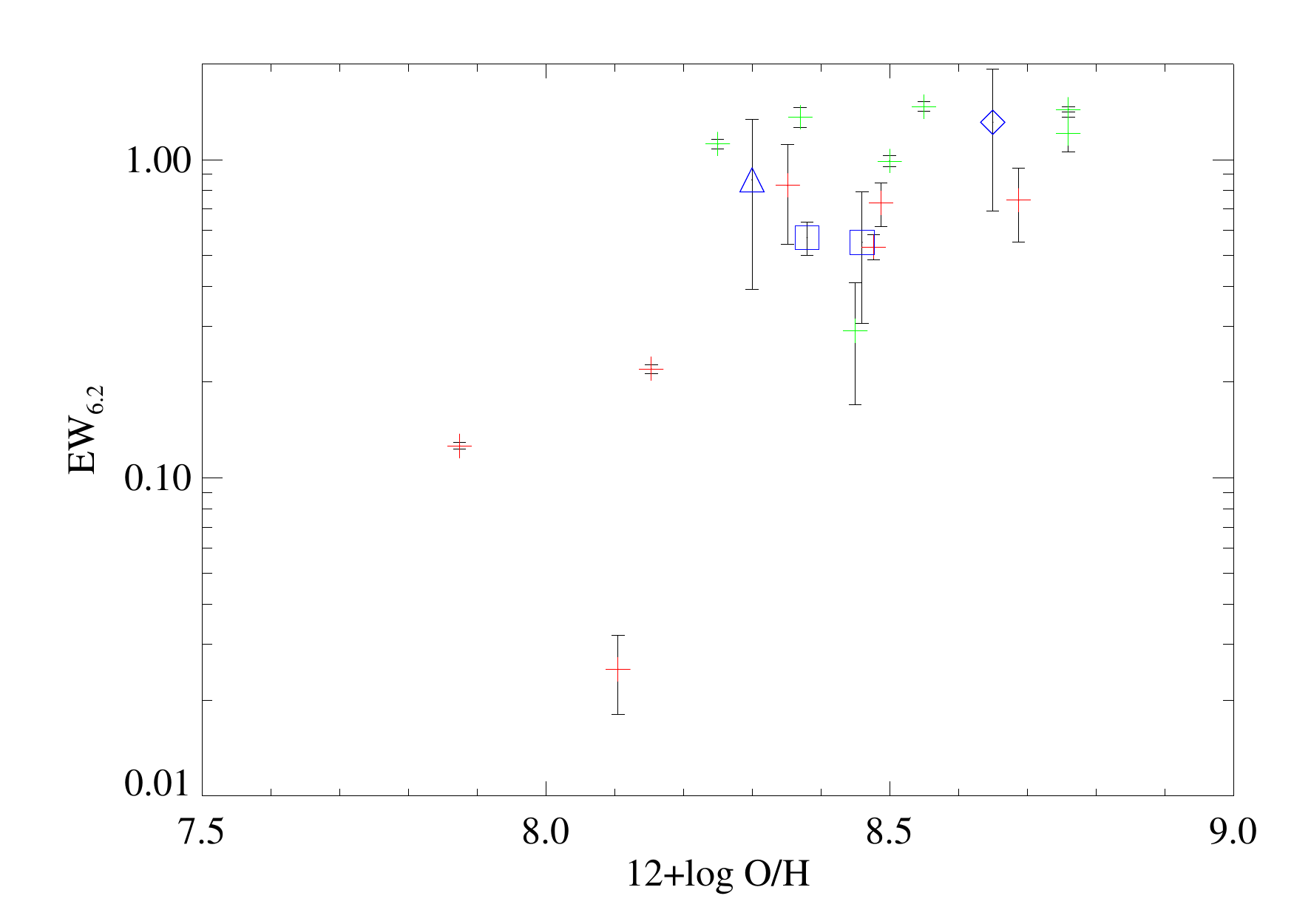}
 \caption{Equivalent width of the 6.2 $\mu$m PAH line versus oxygen abundance in the intergalactic star forming regions observed with the IRS (blue triangle for Arp~105S, diamond for Arp~245N and squares for NGC~5291). The BCDGs from the \citet{wu2006a} sample (red crosses), and the individual star forming regions in M101 (\citet{gordon2008a}, green crosses) are plotted as well.\label{fig:correl-wu-62}}
\end{figure}

\begin{figure}
 \includegraphics[width=\columnwidth]{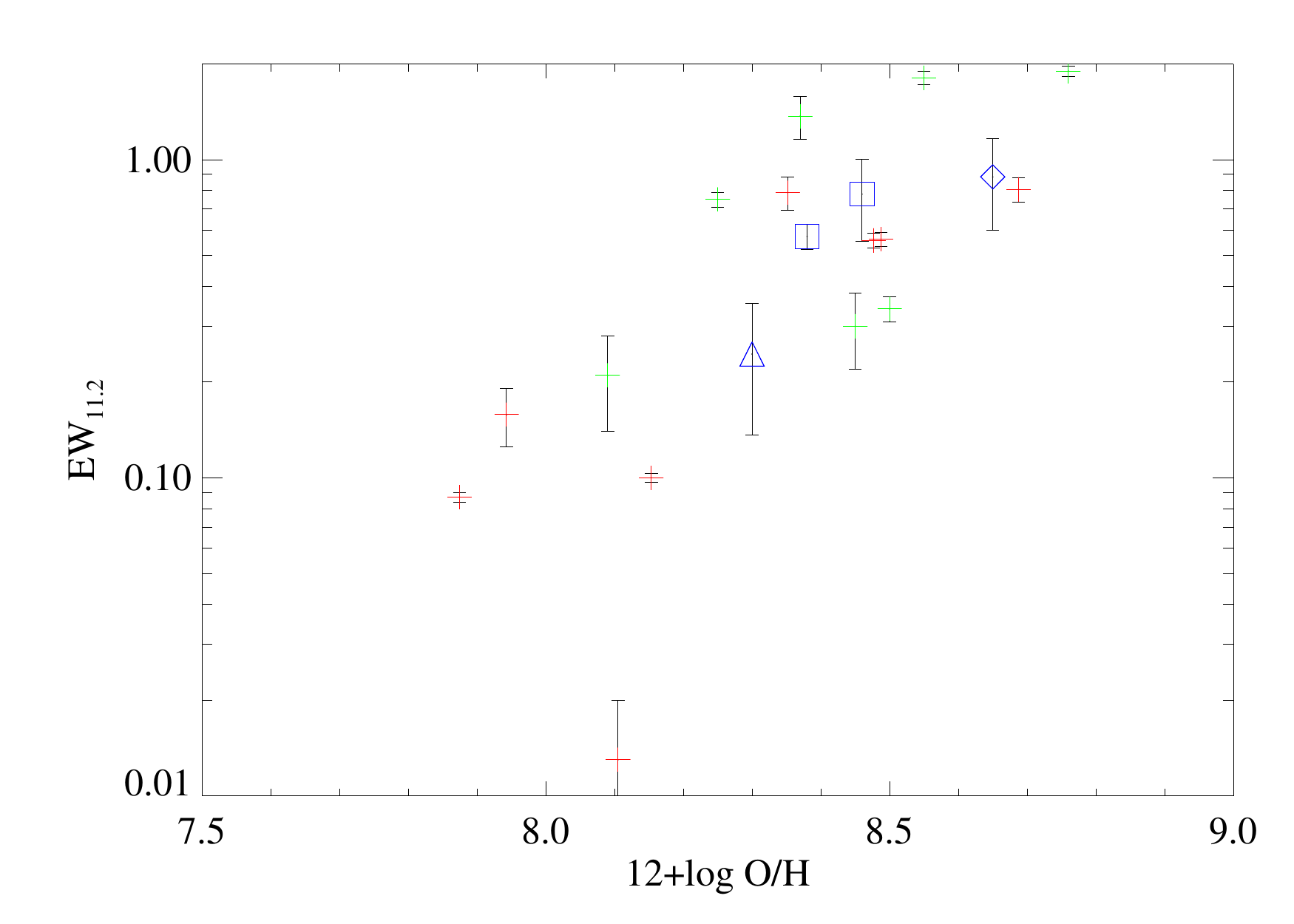}
 \caption{Equivalent width of the 11.2 $\mu$m PAH line versus oxygen abundance in the intergalactic star forming regions observed with the IRS. The color and shape of the symbols used is the same as in Fig.~\ref{fig:correl-wu-62}. \label{fig:correl-wu-11}}
\end{figure}

PAH bands in collisional debris have been directly detected for the first time in the NGC~5291 system \citep{higdon2006a}. The new mid-infrared spectra of the Tidal Dwarf Galaxy candidates in Arp~105S and Arp~245N presented in this paper also exhibit strong PAH emission (see Fig.~\ref{fig:arp105-IRS} and \ref{fig:arp245-IRS}), a property which may result from their relatively high metallicity. It has been established \citep[e.g.,][references there in]{engelbracht2005a,walter2007a} that the PAH emission is correlated with metallicity. In Fig. \ref{fig:correl-wu-62} (respectively \ref{fig:correl-wu-11}), we plot the 6.2 $\mu$m (resp. 11.2 $\mu$m) band equivalent width of the PAH features versus oxygen abundance. Galaxies from the sample of \citet{wu2006a} and individual star-forming regions in the spiral M101 \citep{gordon2008a} are plotted.

These plots show a clear correlation between the strength of the PAH emission and metallicity for the BCDGs and the M101 individual star forming regions. The four intergalactic star forming regions exhibit an aromatic emission at the expected level. Indeed, at a given metallicity, the equivalent widths of the PAH bands emitted by the intergalactic star forming regions are very similar to those of star forming regions within galaxies.

\begin{figure}
 \includegraphics[width=\columnwidth]{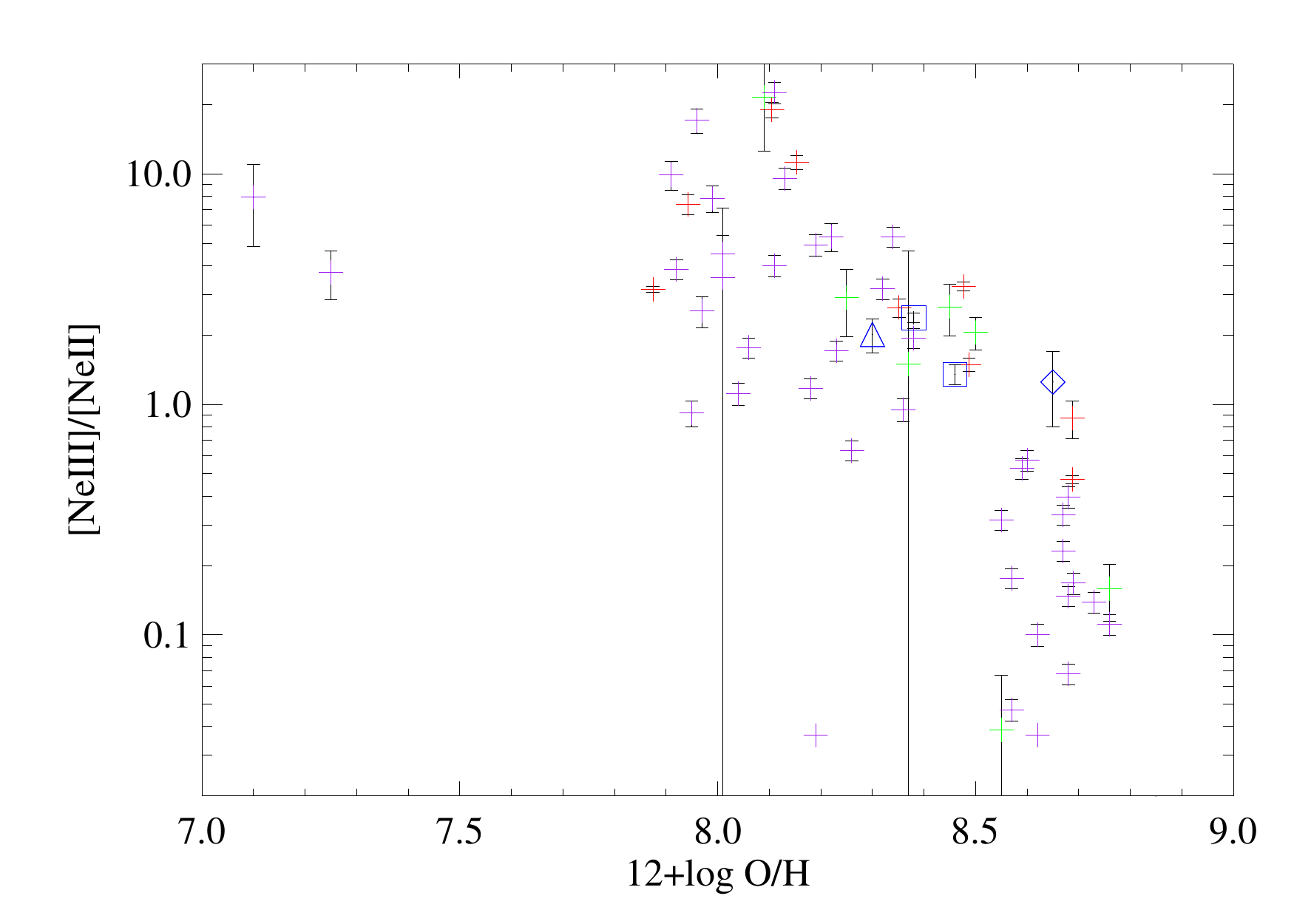}
 \caption{[NeIII]/[NeII], a tracer of the hardness of the radiation field, versus the oxygen abundances. Galaxies from the BCDG sample of \citet{wu2006a} (red crosses), from the starburst sample of \citet{engelbracht2008a} (purple crosses); individual star forming regions in M101 (green crosses), and the intergalactic star forming regions (blue triangle for Arp~105S, diamond for Arp~245N and squares for NGC~5291).\label{fig:correl-wu-z-ne}}
\end{figure}

As shown for instance by \citet{madden2006a}, the strength of the PAH features is also closely connected with the hardness of the radiation field. \citet{gordon2008a} recently claimed that this parameter, itself correlated with metallicity (see Fig.~\ref{fig:correl-wu-z-ne}), is the main parameter governing the evolution of the strength of the PAH features. Indeed, lower metallicity stars emit harder radiation due to the lack of metals in their atmosphere. As a result, the hard radiation field combined with a high luminosity density can destroy PAH carriers \citep{wu2006a}.

\begin{figure}
 \includegraphics[width=\columnwidth]{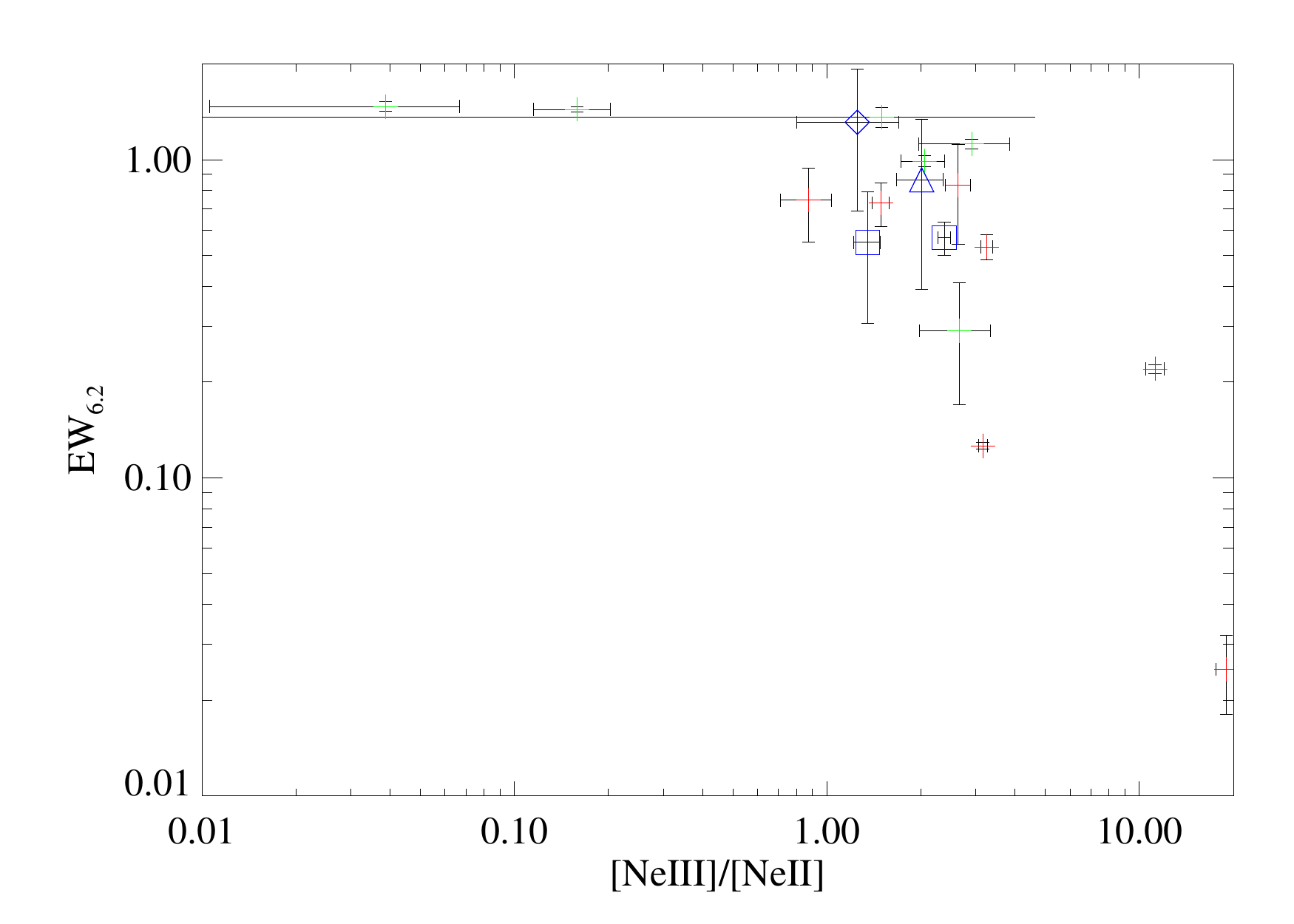}
 \caption{Equivalent width of the PAH 6.2 $\mu$m band as a function of the [NeIII]/[NeII] ratio. The color and shape of the symbols used is the same as in Fig.~\ref{fig:correl-wu-62}.\label{fig:6.2-ionic}}
\end{figure}

\begin{figure}
 \includegraphics[width=\columnwidth]{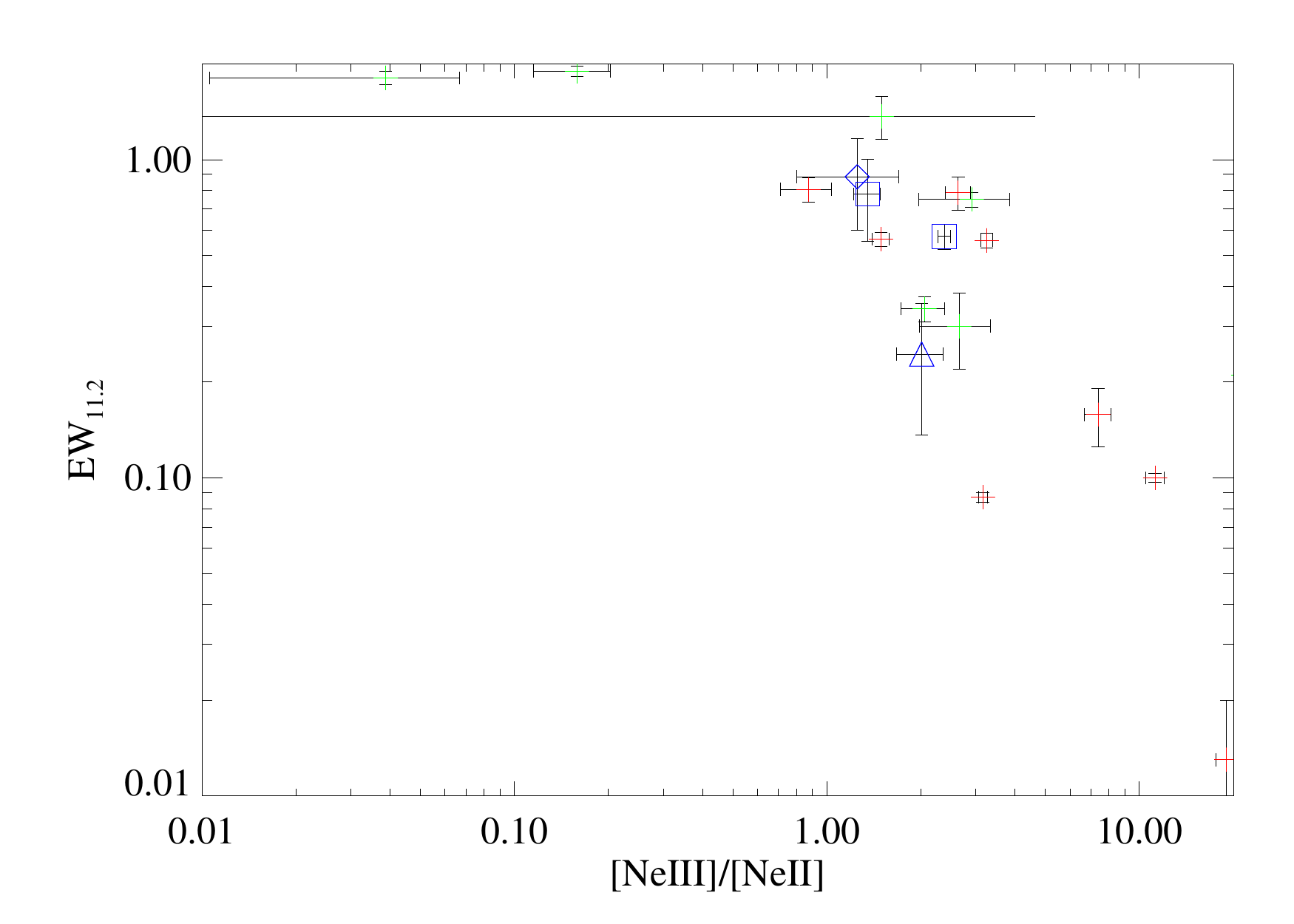}
 \caption{Equivalent width of the PAH 11.2 $\mu$m band as a function of the [NeIII]/[NeII] ratio. The color and shape of the symbols used is the same as in Fig.~\ref{fig:correl-wu-62}.\label{fig:11-ionic}}
\end{figure}

Since we also have high resolution IRS spectra for most regions in our sample we examine in Fig. \ref{fig:6.2-ionic} and \ref{fig:11-ionic} the correlation between the neon line ratio and the EW of the 6.2~$\mu$m and 11.2~$\mu$m features. We find that the regions in the collisional debris fall broadly within the trend discussed by, e.g. \citet{wu2006a}.

Unfortunately mid-infrared spectra are available for only 4 TDGs. For the other collisional debris, we have used the emission in the IRAC 8.0 $\mu$m band as an indirect tracer. The validity of this approach has been discussed in more detail by several authors \citep[see e.g.,][]{engelbracht2005a,engelbracht2008a} and is corroborated by our current results. In our IRS sample, PAH features dominate the emission in this broad band. Continuum emission from hot dust may, however, have a significant contribution which should be taken into account. To that purpose, we have adopted the method presented in \citet{engelbracht2005a,engelbracht2008a}: the 8.0 $\mu$m emission is normalized to the underlying hot dust continuum as follows: $F_{8.0}/\left(F_{4.5}-\alpha F_{3.6}\right)$. This assumes that the 3.6 $\mu$m emission is purely stellar\footnote{In fact this band is also polluted by the 3.3 $\mu$m PAH feature as we mentioned earlier. Unfortunately, its strength may not be directly determined, even in the TDGs with mid-IR spectroscopic data: the IRS instrument does not cover this short wavelength range.}, and that the IRAC 4.5 $\mu$m band traces the hot dust, once the stellar continuum has been subtracted. The value of $\alpha$ -- the conversion factor from the 3.6 $\mu$m to the 4.5 $\mu$m stellar flux -- is given by a spectral synthesis code and depends on the metallicity. For those intergalactic HII regions for which no oxygen abundance has yet been determined, we adopt $\alpha=0.535$, the mean of the coefficients for the $8.15\le Z<8.50$ and $8.50\le Z<8.85$ bins from \citet{engelbracht2008a}.

\begin{figure}
 \includegraphics[width=\columnwidth]{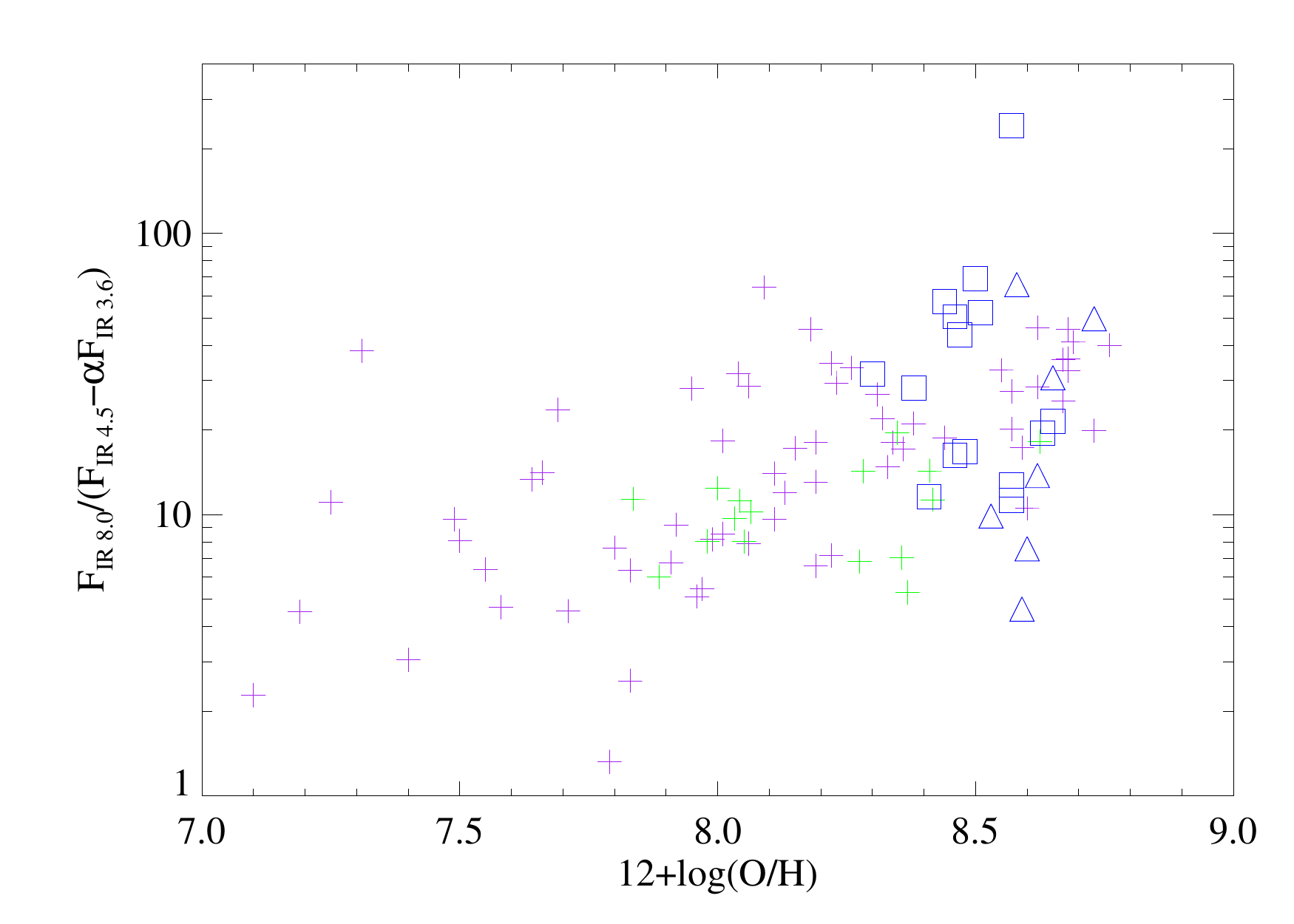}
 \caption{Ratio of the mid-infrared fluxes $F_{8.0}/\left(F_{4.5}-\alpha F_{3.6}\right)$ as a function of the oxygen abundance. Intergalactic star forming regions are sorted in 2 categories whether they contain a significant old stellar population (blue triangles) or they are located within mainly gaseous streams (blue squares). Galaxies in the reference samples: \citet{rosenberg2006a} (green plusses), \citet{engelbracht2008a} (purple plusses).\label{fig:correl-z-pah}}
\end{figure}

\begin{figure}
 \includegraphics[width=\columnwidth]{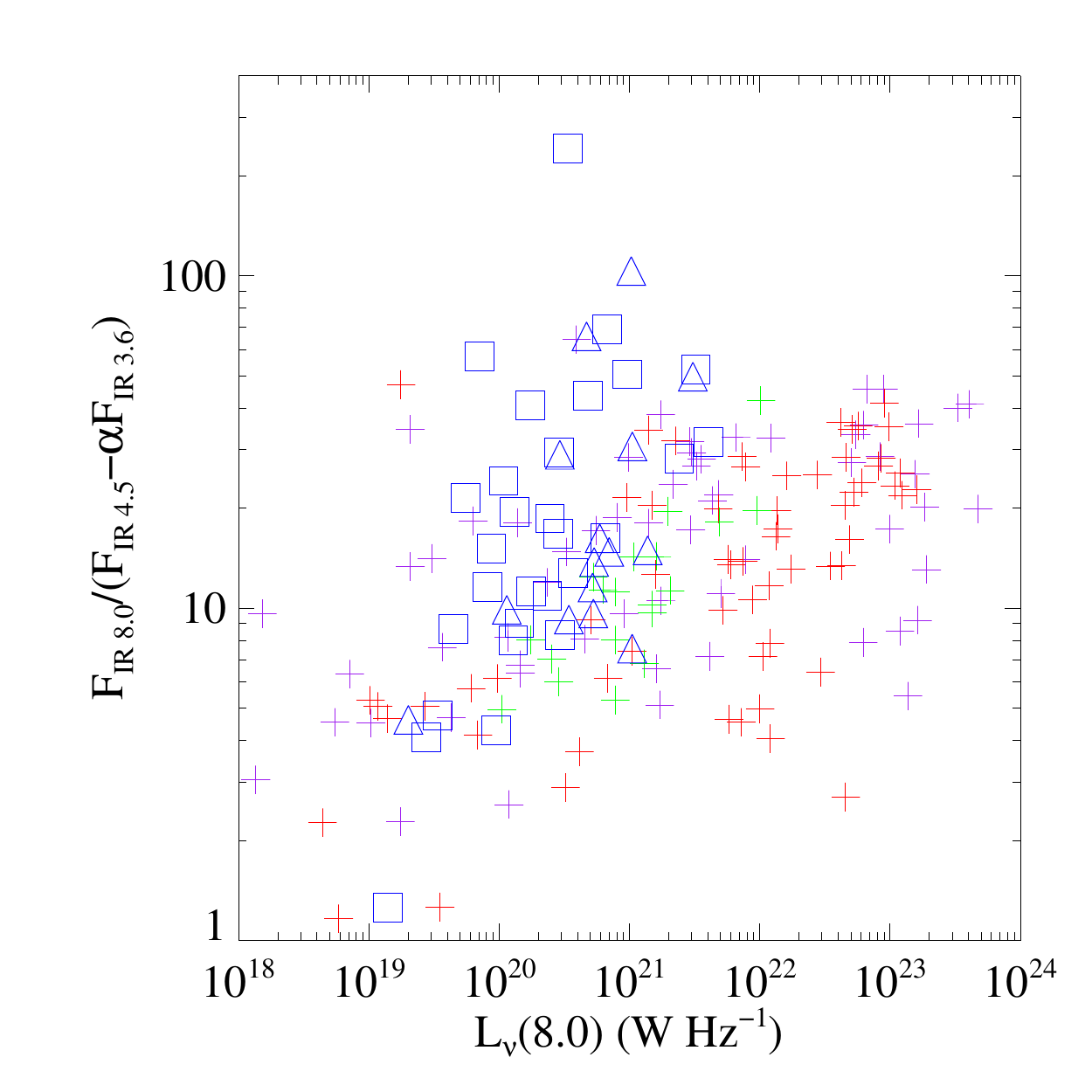}
 \caption{PAH fluxes traced by the 8.0 $\mu$m emission normalized to the underlying hot dust continuum $F_{8.0}/\left(F_{4.5}-\alpha F_{3.6}\right)$ as a function of the 8.0 $\mu$ luminosity. Intergalactic star forming regions are sorted in 2 categories whether they contain a significant old stellar population (blue triangles) or they are located within mainly gaseous streams (blue squares). Galaxies in the reference samples: SINGS (red plusses), \citet{rosenberg2006a} (green plusses) and \citet{engelbracht2008a} (purple plusses).\label{fig:pah-galaxies}}
\end{figure}

\begin{figure}
 \includegraphics[width=\columnwidth]{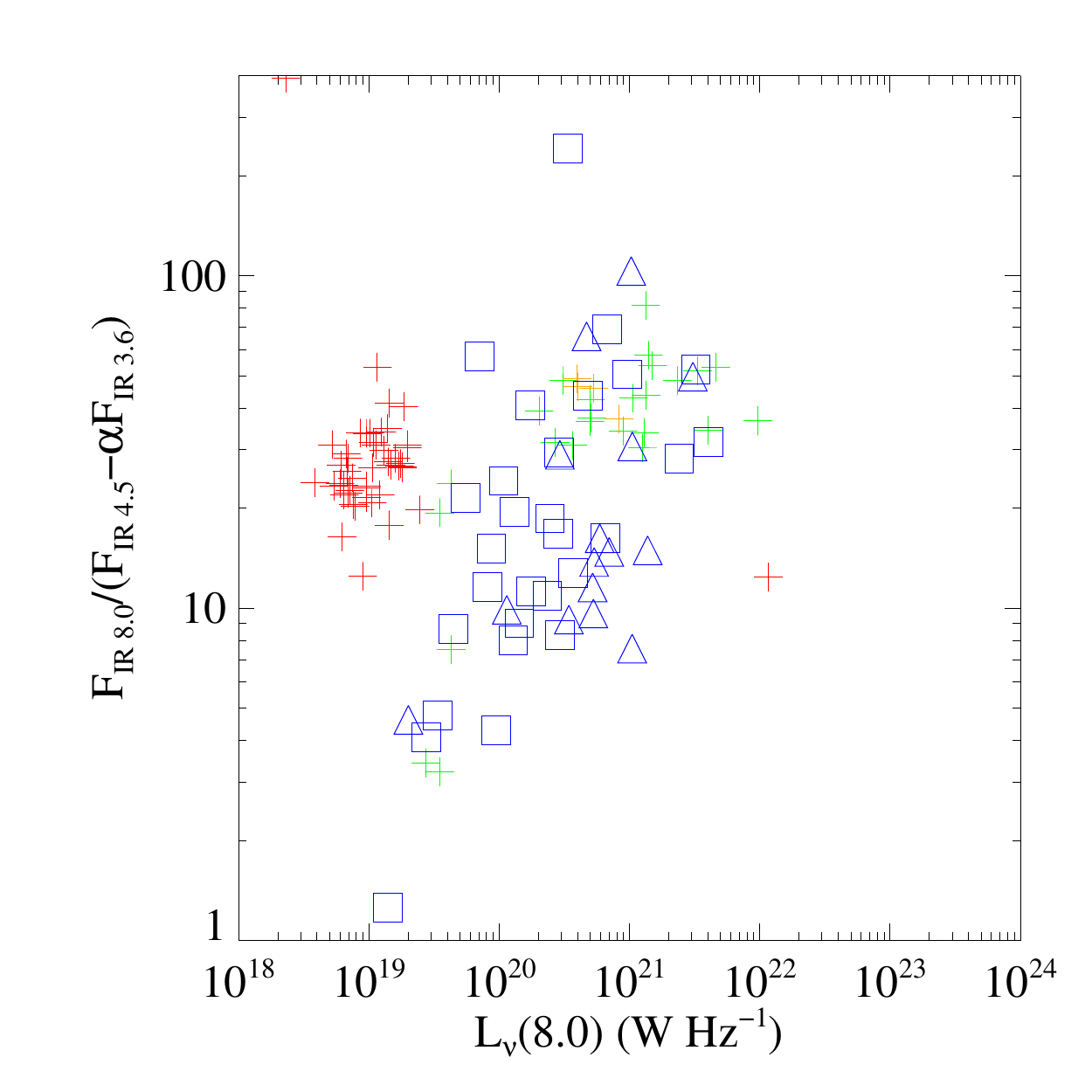}
 \caption{PAH fluxes traced by the 8.0 $\mu$m emission normalized to the underlying hot dust continuum $F_{8.0}/\left(F_{4.5}-\alpha F_{3.6}\right)$ as a function of the 8.0 $\mu$m luminosity. Intergalactic star forming regions are sorted in 2 categories whether they contain a significant old stellar population (blue triangles) or they are located within mainly gaseous streams (blue squares). Galactic star forming regions in the reference samples: M81 (red plusses), Arp~82 (green plusses) and Arp~24 (orange plusses).\label{fig:pah-hii-regions}}
\end{figure}

In Fig. \ref{fig:correl-z-pah} we plot the ratio of the mid-infrared fluxes $F_{8.0}/\left(F_{4.5}-\alpha F_{3.6}\right)$ as a function of oxygen abundance for intergalactic star forming regions and the \citet{rosenberg2006a} dwarf galaxies and the \citet{engelbracht2008a} starburst galaxies comparison samples. The aim is to see if intergalactic star forming regions follow the same well-known relation between the strength of the aromatic emission and metallicity. This plot confirms that intergalactic star forming regions lie along the correlation between the mid-IR emission, corrected for the hot dust and stellar contribution, and metallicity. The intergalactic star forming regions seem to follow the same trend as spiral galaxies, although the dispersion might be slightly higher. This may also be seen in Fig. \ref{fig:pah-galaxies} and \ref{fig:pah-hii-regions} showing the $F_{8.0}/\left(F_{4.5}-\alpha F_{3.6}\right)$ PAH indicator as a function of the 8.0 $\mu$m luminosity for the reference samples. The intergalactic star forming regions describe approximately the upper envelope of the reference sample of integrated galaxies but the mean ratio is similar.

From this study of the dust properties of collisional debris, we conclude that their mid-infrared emission, dominated by PAH, is ``normal'' and comparable to that of the star-forming regions in regular galaxies that have the same metallicity. PAH are emitted in photodissociation regions associated with star-formation episodes and are thus considered as a reliable tracer of star-formation \citep[except in a low-Z environment ; ][]{calzetti2007a}. Thus the calibrations between the IRAC 8.0 $\mu$m band and the SFR, as determined for regular spiral galaxies, should also be valid for collisional debris. This assessment is exploited in the rest of the paper.

\subsection{Comparing Star Formation Rates\label{ssec:SFR}}

We compare here the emission of the three star formation tracers used in this study: 

\begin{itemize}
 \item ultraviolet emission, due to the direct radiation of mostly massive stars,
 \item H$\alpha$ emission from regions ionized by young, very massive stars,
 \item mid-infrared 8.0~$\mu$m emission due to the PAHs in photodissociation regions.
 \end{itemize}

These tracers, although all somehow related to the star-formation process, have their own characteristics and biases, which are discussed in section \ref{ssec:env-effects}.

To put the UV, H$\alpha$ and MIR luminosities on a common scale, we have converted them into units corresponding to measures of the SFR, using the following formulas:

\begin{itemize}
 \item SFR(UV)=$1.4\times10^{-21}L_\nu$ [W~Hz$^{-1}$]
 \item SFR(H$\alpha$)=$7.9\times10^{-35}L(H\alpha)$ [W]
 \item SFR(8.0 $\mu$m)=$1.88\times10^{-36}\nu L_\nu$ [W~Hz$^{-1}$]
\end{itemize}

These conversion factors are based on the values determined by \citet{kennicutt1998a} for the UV and H$\alpha$ luminosities, and \citet{wu2005a} for the 8.0~$\mu$m emission. One should note that they are only valid for specific star formation histories, in particular a star formation rate which is assumed to be constant for 100 Myr and with a Salpeter IMF. As discussed in sect.~\ref{ssec:stellar-pops}, these conditions may not apply to collisional debris, and thus the derived SFR may be off by some unknown factor, especially in the youngest regions. This scaling is done here mainly to facilitate comparison between the three tracers (i.e. their fluxes) and to compare the SFR in debris with that of other star-forming regions.

\begin{figure}
 \includegraphics[width=\columnwidth]{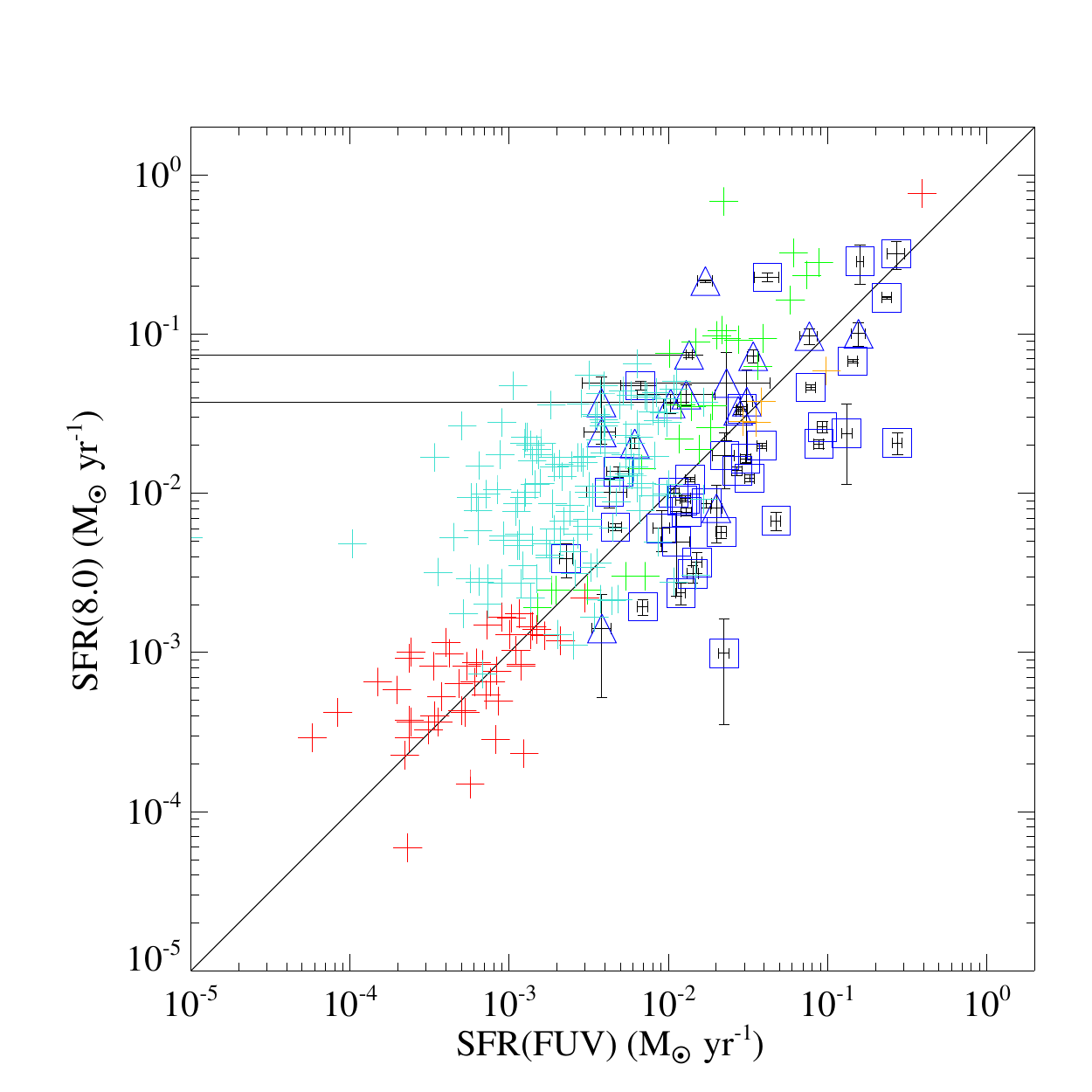}
 \caption{Comparison of the foreground Galactic extinction corrected SFR estimated from FUV and 8.0 $\mu$m. Intergalactic star forming regions are sorted in 2 categories whether they contain a significant old stellar population (blue triangles) or they are located within mainly gaseous streams (blue squares). Galaxies in the reference samples: M~51 (turquoise plusses), M~81 (red plusses) and Arp~82 (green plusses). The black solid line indicates when the FUV derived and the 8.0~$\mu$m star formation rates are identical.\label{fig:sfr-fuv-ir80-hii-regions}}
\end{figure}

\begin{figure}
 \includegraphics[width=\columnwidth]{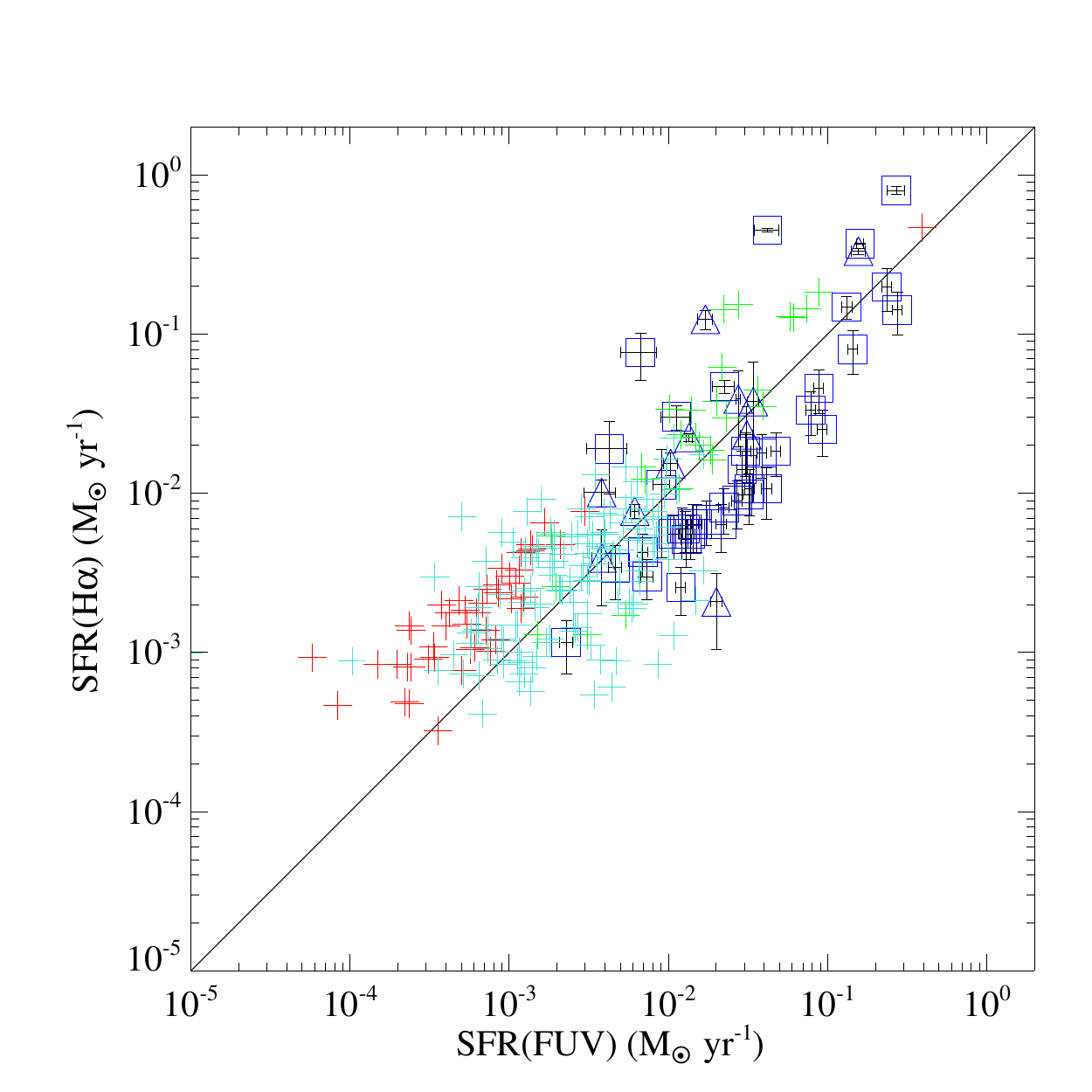}
 \caption{Comparison of the foreground Galactic extinction corrected SFR estimated from FUV and H$\alpha$. The color and shape of the symbols used is the same as in Fig.~\ref{fig:sfr-fuv-ir80-hii-regions}.\label{fig:sfr-fuv-halpha-hii-regions}}
\end{figure}

In Fig. \ref{fig:sfr-fuv-ir80-hii-regions} (resp. \ref{fig:sfr-fuv-halpha-hii-regions}) we compare the SFR (uncorrected for extinction) estimated from the FUV band with that estimated from the 8.0 $\mu$m (resp. H$\alpha$). \citet{boquien2007a} have shown that the intergalactic SF regions around NGC~5291 exhibit a strong, systematic, ultraviolet excess that can be clearly seen in figure \ref{fig:sfr-fuv-halpha-hii-regions} where H$\alpha$ appears depressed compared to the FUV emission. On average, this is not the case in the other interacting galaxies in our sample (see Tables \ref{tab:comp-sfr} and \ref{tab:mean-ratios}). This UV excess is found only locally, in specific star-forming regions, which mostly are hosted by gas-rich collisional debris. The dispersion among different regions is important, but not dramatically larger than that measured in star-forming regions belonging to a single galaxy, like M51. The 8.0 $\mu$m emission is itself the strongest for regions which contain an old stellar population. This is probably due to the fact that older stars can also heat the dust and therefore create diffuse emission. This kind of emission is particularly visible in spiral galaxies and can account for a significant fraction of the 8.0~$\mu$m flux as suggested by \citet{calzetti2007a}.

\begin{table*}[!htbp]
\begin{minipage}[c]{\textwidth}
\centering
\caption{Total star formation rate in the selected star forming regions, therefore excluding parent galaxies, according to the three tracers: far ultraviolet, H$\alpha$ and infrared at 8.0 $\mu$m.\label{tab:comp-sfr}}
\begin{tabular}{c c c c}
\hline\hline
Sample&SFR(FUV)&SFR(H$\alpha$)&SFR($8.0$)\\
&(M$_\sun$\,yr$^{-1}$)&(M$_\sun$\,yr$^{-1}$)&(M$_\sun$\,yr$^{-1}$)\\\hline
Stephan's Quintet&$0.71\pm0.08$&$>2.08\pm0.10$&$1.06\pm0.14$\\
Arp~105&$0.28\pm0.08$&$0.37\pm0.002$&$0.59\pm0.16$\\
Arp~245&$0.03\pm0.002$&$0.05\pm0.006$&$0.15\pm0.01$\\
NGC~5291&$1.44\pm0.10$&$>0.71\pm0.22$&$>0.54\pm0.02$\\
NGC~7252&$0.08\pm0.01$&$0.07\pm0.03$&$0.08\pm0.03$\\
VCC~2062&$0.002\pm0.0002$&$0.001\pm0.0004$&$0.004\pm0.0009$\\\hline
Rosenberg&--&--&$0.14$\\
SINGS&$0.41$&--&$1.86$\\\hline
M~51&$0.48$&$0.54$&$2.08$\\
M~81&$0.39$&$0.47$&$0.77$\\
Arp~82&$0.61$&$1.31$&$2.60$\\
Arp~24&$0.20$&--&$0.15$\\
\hline
\end{tabular}
\tablenotetext{}{In the case of the \citet{rosenberg2006a} and SINGS samples, it is the star formation rate integrated over the entire object.}
\tablenotetext{}{The star formation rates of the reference samples are computed with the same conversion factor as for the intergalactic star forming regions, using the published luminosities.}
\end{minipage}
\end{table*}

\begin{table*}[!htbp]
\begin{minipage}[c]{\textwidth}
\centering
\caption{Mean ratio between the PAH emission and the dust continuum; mean ratio between the SFR estimated from the 8.0 $\mu$m emission and from the far ultraviolet emission and mean ratio of the SFR estimated from the H$\alpha$ emission and from the far ultraviolet emission.\label{tab:mean-ratios}}
\begin{tabular}{c c c c}
\hline\hline
Sample&$\left<\textrm{F}_{8.0}/\left(\textrm{F}_{4.5}-\alpha\textrm{F}_{3.6}\right)\right>$&$\left<\textrm{SFR}\left(8.0\right)/\textrm{SFR}\left(\textrm{FUV}\right)\right>$&$\left<\textrm{SFR}\left(H\alpha\right)/\textrm{SFR}\left(\textrm{FUV}\right)\right>$\\
\hline
Stephan's Quintet&$84\pm74$&$3.0\pm3.7$&$4.3\pm3.8$\\
Arp~105&$16\pm8$&$3.6\pm3.5$&--\\
Arp~245&$20\pm11$&$4.7\pm1.5$&$1.8\pm0.6$\\
NGC~5291&$17\pm14$&$0.5\pm0.3$&$0.4\pm0.1$\\
NGC~7252&$24\pm28$&$0.8\pm0.5$&$0.8\pm0.6$\\
VCC~2062&$21$&$1.7$&$0.5$\\\hline
debris with older stars&$26.0\pm27.3$&$3.6\pm3.6$&$1.9\pm1.9$\\
debris without older stars&$29.9\pm46.7$&$1.0\pm1.5$&$1.4\pm2.5$\\
Rosenberg&$13\pm8$&--&--\\
SINGS&$16\pm11$&$9.6\pm26.8$&--\\
Engelbracht&$20\pm11$&--&--\\
\hline
M~51&--&$7.2\pm9.0$&$1.7\pm1.8$\\
M~81&$33\pm48$&$1.4\pm1.1$&$3.2\pm2.0$\\
Arp~82&$36\pm17$&$3.8\pm5.7$&$2.0\pm1.4$\\
Arp~24&$45\pm5$&$0.8\pm0.2$&--\\
\hline
\end{tabular}
\tablenotetext{}{Aberrant points are removed before calculating the above values. Those were due to regions having a very low $\left(\textrm{F}_{4.5}-\alpha\textrm{F}_{3.6}\right)$ flux.}
\end{minipage}
\end{table*}

\section{Discussion\label{sec:discussion}}

\subsection{The SFR in tidal debris}

The objects in our primary list were selected for having prominent star-forming regions in their collisional debris. This may be quantified using the ultraviolet maps, a reliable tracer of recent star-formation episodes. Table \ref{tab:frac-sfr-igm} lists the UV fluxes, converted to M$_\sun$\,yr$^{-1}$ (using the relation given in section~\ref{ssec:SFR}), from regions located within the optical radius of the parent galaxies and outside it, within the collisional debris. The instantaneous fraction of stars born in the ``intergalactic'' medium is as high as 80\% with a median value of nealy 21\%. The star-formation rate in this environment may exceed 1 M$_\sun$\,yr$^{-1}$, a value which corresponds to the measure of the SFR integrated over the entire disk of normal spiral galaxies! This star formation is rate is a lower bound as it is not corrected for internal extinction\footnote{Spectroscopic observations necessitate long exposure times to observe all the selected star forming regions in all the systems. In addition the use of the HI column density to evaluate the extinction is impaired by the lack of high resolution observations for some of the systems.}. Even if the extinction may be higher in spiral galaxies than in intergalactic star forming regions, \citet{boquien2007a} showed that using the 8.0~$\mu$m indicator, the intergalactic star formation rate still accounted for over 50\% of the total star formation rate in NGC~5291.

For comparison, in less active (and more typical) nearby interacting systems, the tidal structures account for 10\% at most of the total star formation rate \citep{struck2007a,smith2007a,hancock2007a}. Their result, uncorrected for extinction, holds whether the systems are observed in infrared, H$\alpha$ or ultraviolet.

\begin{table*}[!htbp]
\centering
\caption{Star formation rate in the intergalactic medium, total star formation rate in the system and fraction of stars currently formed in the intergalactic medium.\label{tab:frac-sfr-igm}}
\begin{tabular}{c c c c }
\hline\hline
System&SFR$_{IGM}$(FUV)&SFR$_{total}$(FUV)&SFR$_{IGM}$(FUV)/SFR$_{total}$(FUV)\\
&(M$_\sun$\,yr$^{-1}$)&(M$_\sun$\,yr$^{-1}$)&\\\hline
Stephan's Quintet&0.71&3.41&20.8\\
Arp~105&0.28&0.64&43.2\\
Arp~245&0.03&1.33&2.5\\
NGC~5291&1.44&1.69&85.1\\
NGC~7252&0.08&0.73&11.3\\
VCC~2062&0.002&0.043&4.6\\
\hline
\end{tabular}
\end{table*}

One of the main characteristics of the collisional debris studied here is their high gas content, reflecting the fact that their parent galaxies were gas rich. This situation is rather rare in the local Universe but was much more common at high redshift when galaxies had higher gas fractions.

\subsection{Factors affecting the inferred SFR\label{ssec:env-effects}}

This paper makes use of three SFR diagnostics, which are sensitive to a number of factors:

\begin{itemize}
 \item timescale: H$\alpha$ is sensitive to the most recent SF (nearly instantaneous compared to ultraviolet), giving an estimate of the SFR averaged over a few million years. On the other hand ultraviolet emission gives an estimate of the SFR averaged over about 100 Myr,
 \item metallicity: the relation between PAH luminosity and star formation rate is not linear and depends on the metallicity. The H$\alpha$ luminosity is also metal-dependent, as shown by \citet{bicker2005a},
 \item extinction: ultraviolet radiation suffers the most from extinction ($A_{FUV}/A_{H\alpha}\sim3.1$),
 \item geometric factors: H$\alpha$ is sensitive to the effects of geometry. Young stars are enshrouded in dust clouds and their ionizing radiation is efficiently absorbed by dust,
 \item variations of the initial mass function: depending on the distribution of the mass of stars at birth, the amount of energetic radiation emitted per unit mass changes. Since UV emission is not only due to the most massive stars, changes in the shape of the IMF will affect differently the UV and H$\alpha$ luminosities.
\end{itemize}

In principle, ages, extinction or even the IMF of a region of star formation may be constrained by measuring flux ratios of UV/$\left[8.0\right]$ or UV/H$\alpha$, when coupled with evolutionary synthesis models. In practice, strong degeneracies make this exercise particularly delicate.

The aim here is to find which factors can explain the apparent diversity of intergalactic star forming regions. This is particularly important to better understand the star formation process in tidal debris and find what parameter is the driving factor. In this section we verify the influence of the extinction, metallicity, or stellar populations. We have made here such a comparison between the various star-formation indicators for about 60 individual star-forming regions in collisional debris. We discuss below the effect of each parameter on the measure of the SFR, or more precisely on the scatter of the various estimates from one region to the other (see Section~\ref{ssec:SFR}); we will try to determine which parameter best manages to reduce it.

\subsubsection{Extinction}

To study the influence of extinction, we corrected the star formation rates using the \citet{calzetti2000a} extinction curve (A$_\textrm{FUV}$/A$_\textrm{V}=2.54$ and A$_{\textrm{H}\alpha}$/A$_\textrm{V}=0.82$). The optical extinction was derived from the Balmer decrement\footnote{NGC~7252 and Stephan Quintet are not published yet. For Arp~105, Arp~245, NGC~5291 and VCC~2062 refer to \citet{duc1994a,duc2000a,duc1998a,duc2007a}}, as measured from spectra of the HII regions. No correction was applied in the mid-infrared regime. The results are presented in Table \ref{tab:evol-disp-extinc}.

\begin{table*}[!htbp]
\tiny
\centering
\caption{Ratio of the star formation rates $\Delta\left(\textrm{SFR}\left(8.0\right)/\textrm{SFR}\left(\textrm{FUV}\right)\right)/\left<\textrm{SFR}\left(8.0\right)/\textrm{SFR}\left(\textrm{FUV}\right)\right>$ and $\Delta\left(\textrm{SFR}\left(\textrm{H}\alpha\right)/\textrm{SFR}\left(\textrm{FUV}\right)\right)/\left<\textrm{SFR}\left(\textrm{H}\alpha\right)/\textrm{SFR}\left(\textrm{FUV}\right)\right>$ raw and corrected for the extinction.\label{tab:evol-disp-extinc}}
\begin{tabular}{c c c c c c c}
\hline\hline
System&$\left<\mathrm{A_V}\right>$&$\frac{\Delta\mathrm{A_V}}{\left<\mathrm{A_V}\right>}$&$\frac{\Delta\left(\textrm{SFR}\left(8.0\right)/\textrm{SFR}\left(\textrm{FUV}\right)\right)}{\left<\textrm{SFR}\left(8.0\right)/\textrm{SFR}\left(\textrm{FUV}\right)\right>}$&$\left(\frac{\Delta\left(\textrm{SFR}\left(8.0\right)/\textrm{SFR}\left(\textrm{FUV}\right)\right)}{\left<\textrm{SFR}\left(8.0\right)/\textrm{SFR}\left(\textrm{FUV}\right)\right>}\right)_\textrm{corr.}$&$\frac{\Delta\left(\textrm{SFR}\left(\textrm{H}\alpha\right)/\textrm{SFR}\left(\textrm{FUV}\right)\right)}{\left<\textrm{SFR}\left(\textrm{H}\alpha\right)/\textrm{SFR}\left(\textrm{FUV}\right)\right>}$&$\left(\frac{\Delta\left(\textrm{SFR}\left(\textrm{H}\alpha\right)/\textrm{SFR}\left(\textrm{FUV}\right)\right)}{\left<\textrm{SFR}\left(\textrm{H}\alpha\right)/\textrm{SFR}\left(\textrm{FUV}\right)\right>}\right)_\textrm{corr.}$\\\hline
Stephan's Quintet&$1.2$&$0.7$&$1.3$&$1.3$&$0.8$&$0.8$\\
NGC~5291&$0.2$&$0.8$&$0.7$&$1.0$&$0.3$&$0.4$\\
NGC~7252&$1.6$&$0.4$&$0.7$&$1.3$&$1.2$&$1.4$\\\hline
debris with older stars&$1.4$&$0.4$&$0.5$&$0.7$&$0.8$&$1.0$\\
debris without older stars&$0.4$&$1.2$&$1.5$&$1.0$&$2.0$&$0.6$\\\hline
\end{tabular}
\tablenotetext{}{Only regions having a star formation rate of at least 0.02 M$_\sun$\,yr$^{-1}$ have been selected.}
\end{table*}

We see that the scatter in the extinction correction is stable among the intergalactic star forming regions in Stephan's Quintet, but increases in NGC~5291, as already mentioned by \citet{boquien2007a}, and in NGC~7252. On average however, when a significant reduction of the relative scatter is observed (which is the case only for debris without older stars), it is comprised between $\sim$30\% and $\sim$70\%. Given these results, we cannot claim that the extinction is the main parameter governing the discrepancies between the measures of SFR(H$\alpha$), SFR(NUV) and SFR(8.0).

\begin{figure}
 \includegraphics[width=\columnwidth]{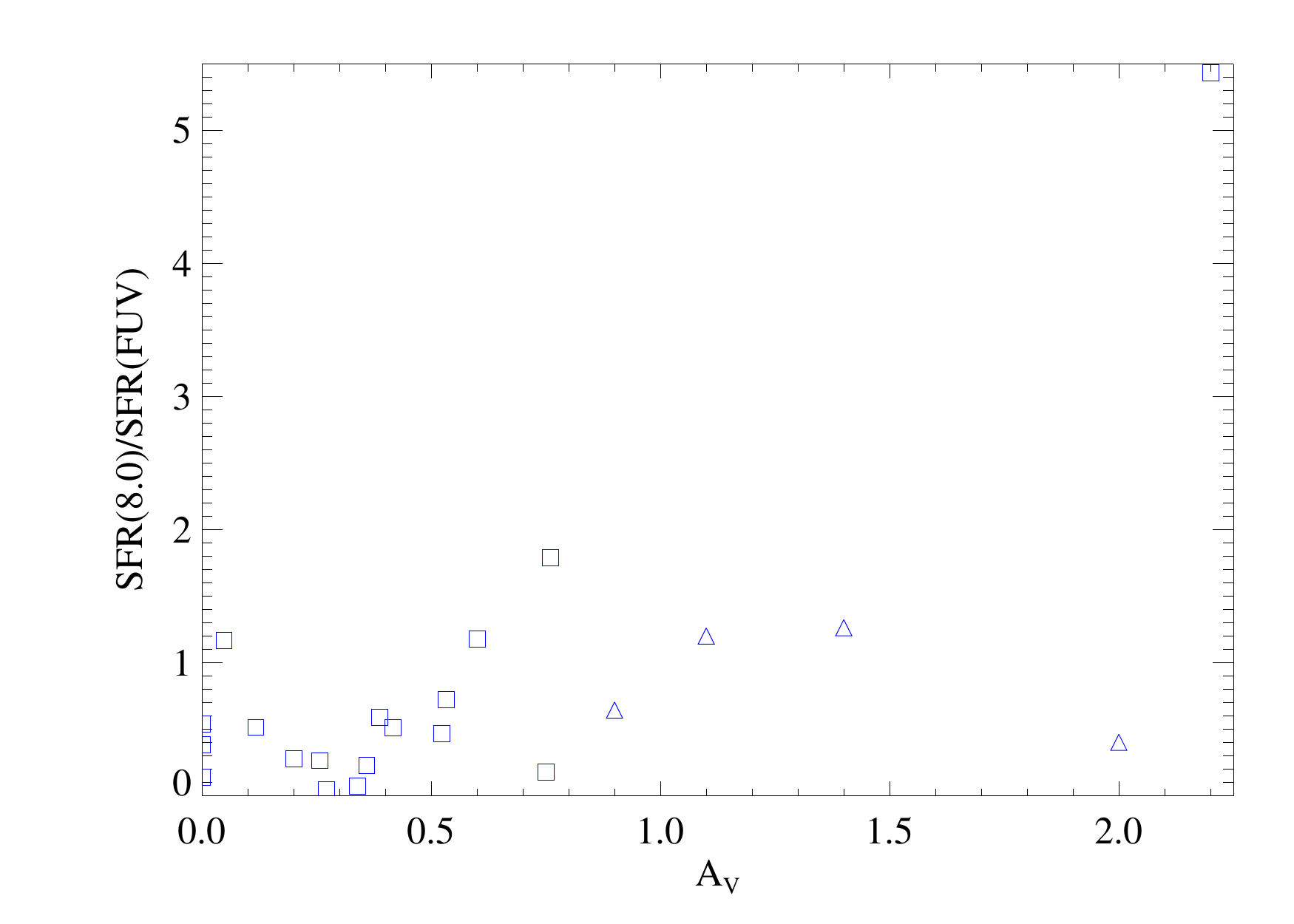}
 \caption{Variation of the SFR(8.0)/SFR(FUV) ratio as a function of A$_\mathrm{V}$. Intergalactic star forming regions are sorted in 2 categories whether they contain a significant old stellar population (blue triangle) or are located within mainly gaseous streams (blue square). Only regions having a star formation rate of at least 0.02 M$_\sun$\,yr$^{-1}$ have been selected. The extinctions have been deduced from spectroscopic observations making use of the Balmer decrement except for NGC~5291 where both the Balmer decrement and HI column density have been used \citep[see][for details]{boquien2007a}.
 \label{fig:av-ir80-ov-fuv}}
\end{figure}

Fig.~\ref{fig:av-ir80-ov-fuv} plots the ratio of the SFR(8.0) over SFR(FUV) as a function of extinction. We are only including regions with equivalent SFR exceeding 0.02 M$_\sun$\,yr$^{-1}$ in order to ensure that the measure of the Balmer decrement, and thus of the extinction, is reliable. The correlation between SFR(8.0)/SFR(FUV) and A$_\mathrm{V}$ is weak. The reduction of the scatter mentioned above is largely due to the regions most affected by extinction, such as SQ-B in Stephan's Quintet, where A$_\mathrm{V} \sim$3.0 magnitudes.

\subsubsection{Metallicity}

The non-linearity of the correlation between the 8.0 $\mu$m flux and metallicity, measured by \citet{calzetti2007a} is largely due to the dependence of the PAH emission on metallicity. The collisional debris shows, however, a rather restricted range of oxygen abundance: 12+log(O/H)=8.3--8.9, close to solar. Why the metallicity in regions located far away from their parent galaxies is on average so high and so uniform remains to be explained. Nevertheless, as far as star formation is concerned, it means that the the IRAC 8.0 $\mu$m traces the SFR linearly. The narrow range of relatively high metallicities should therefore not increase the scatter of the ratio of SFR(8.0)/SFR(FUV).

\subsubsection{Stellar populations\label{ssec:stellar-pops}}

The calibration between the UV/H$\alpha$/MIR luminosities and the star-formation rates, and thus their relative ratio, strongly depends on the age of the most recent starburst as well as on the current and past star formation history. \citet{boquien2007a} have shown that the strong scatter of the UV to H$\alpha$ luminosity ratios, measured along the collisional ring of NGC~5291, and globally its UV excess, were mainly due to age effects: for an instantaneous young starburst, the UV/H$\alpha$ ratio increases quickly with time, so that small delays between the onset of star-formation along the ring will induce rapid changes in the UV/H$\alpha$ ratio.

The relative fraction of young to old stars in a given region may affect the UV to SFR calibration as well, since not only young massive stars emit ultraviolet radiation. Thus variations of the UV to H$\alpha$ ratio may partially reflect contamination of the ultraviolet by a substantially older stellar population. To check this, we have distinguished in all our comparison plots (Fig.~\ref{fig:sfr-fuv-ir80-hii-regions} to \ref{fig:av-ir80-ov-fuv}) collisional debris clearly dominated by old stars pulled out from the parent galaxies (e.g. Arp~105N, Arp~245N) to those apparently formed in pure gaseous tails (Arp~105S, NGC~5291, VCC~2062) and hence made almost exclusively of young stars. The former seem to be UV deficient (but also to have larger extinction) as we can see in Table~\ref{tab:mean-ratios}. This shows that significant ultraviolet contamination by an older stellar population is unlikely.

Finally, the Initial Mass Function will also affect the value deduced for the SFR. A top heavy IMF will boost the emission due to massive stars, and lead to overestimating the total SFR, if measured on the basis of H$\alpha$ luminosity. However as the collision debris are a low density environment such an IMF would be unexpected \citep{krumholz2008a}. It would require full analysis of the Spectral Energy Distribution of tidal debris over a broad wavelength range, and not only the three SF tracers listed here, to draw more definitive conclusions. This is beyond the scope of this paper.

\subsection{Star formation and environment}

As noted before, the discrepancies between the different SFR indicators indicate variations of the local properties of star-forming regions. We measure some scatter from one intergalactic region to another. This is, however, similar to what is observed in individual star-forming regions of a single galactic disk. Besides, the MIR to UV ratio is on average similar in the collisional debris and in spiral disks {\it of the same metallicity}. More generally, despite the differences in the large scale environments, it is remarkable that intergalactic and galactic regions look so similar, at least when studied via the three main SF diagnostics. The presence of strong PAH features in the collisional debris perfectly illustrates this.

\section{Conclusions\label{sec:conclusion}}

We have studied star formation within collisional debris of galaxy-galaxy interactions, including Tidal Dwarf Galaxies and compared it to that occurring in regular spiral galaxies. About 60 star-forming regions in 6 interacting systems showing prominent SF (with integrated current intergalactic SFR that can reach more than one solar mass per year and half of them comprising more than 20\% of that of the whole system) were selected. We used three SF tracers: the UV, measured from PI and archive images from GALEX, H$\alpha$ from ground-based Fabry-Perot and narrow-band images and mid-infrared emission from IRAC on Spitzer. We also present two new spectroscopic observations in the mid-infrared with the Spitzer/IRS instrument in addition to those published by \citet{higdon2006a}. We have obtained the following results:

\begin{enumerate}
 \item Dust emission in tidal debris, as probed with broad band images and spectra is at a level similar to that encountered in star forming regions of comparable metallicity and luminosity in galaxies of a comparison sample.
 \item The star formation rate inferred from ultraviolet, H$\alpha$ and mid-infrared show similar trends and scatter as compared to classical star forming regions in galaxies.
 \item Extinction and metallicity are not the main parameters governing the scatter in the properties of intergalactic star forming regions. Age effects, and variations in the stellar populations, seem to play a dominant role.
\end{enumerate}

The similarity of the star formation properties in intergalactic star forming regions and within spiral galaxies leads us to conclude that star formation processes seem to be mainly driven by local properties -- gas column density, dust content, etc. -- rather than by large-scale environmental factors. Star formation may proceed in intergalactic space the same way as in spiral disks. As a result, intergalactic star forming regions may be used as a laboratory to study star formation: results obtained on intergalactic star forming regions can be applied to star forming regions in galaxies in general, provided the metallicity is similar. The main advantage of the latter environment is its simplicity: the individual star-forming regions are physically isolated, simplifying flux measurements. Moreover, it is worth remembering that the influence of shocks, magnetic fields, etc. is much reduced in the intergalactic medium.

The absence of an old stellar component in at least a fraction of collisional debris makes any analysis of their Spectral Energy Distribution less degenerate, giving hope to constrain key parameters, such as the Initial Mass Function. A variation of the IMF in extreme environments is actively discussed in the literature: in the outermost low density star forming regions in spirals \citep{krumholz2008a,pflamm2008a} as well as in ultra compact dwarf galaxies \citep{mieske2008a}. Other strongly debated questions, such as the threshold for star formation in gas clouds and the star formation efficiency, may be addressed using collisional debris. We are presently collecting a large photometric database to investigate these questions in tidal dwarf galaxies. They will be addressed in future papers.

\begin{acknowledgements}
This research has made use of the NASA/IPAC Extragalactic Database (NED) which is operated by the Jet Propulsion Laboratory, California Institute of Technology, under contract with the National Aeronautics and Space Administration.

GALEX is a NASA Small Explorer, launched in April 2003. We gratefully acknowledge NASA's support of the construction, operation, and science analysis for the GALEX mission, developed in cooperation with the Centre national d'\'etudes spatiales, France, and the Korean Ministry of Science and Technology.

Based on observations obtained at the Canada-France-Hawaii Telescope (CFHT) which is operated by the National Research Council of Canada, the Institut National des Sciences de l'Univers of the Centre National de la Recherche Scientifique of France, and the University of Hawaii.

MB acknowledges support by NASA-ADP grant NNX07AN90G and D. Calzetti for fruitful discussions. UL acknowledges financial support by the Spanish Science Ministry under grant AYA2007-67625-C02-02 and by the Junta de Andaluc\'ia. VC would like to acknowledge partial support from EU ToK grant 39965. Finally we also thank the referee for useful comments.
\end{acknowledgements}

\acknowledgments

{\it Facilities:} \facility{CAO:2.2m}, \facility{CFHT}, \facility{ESO:3.6m}, \facility{GALEX}, \facility{KPNO:2.1m}, \facility{NTT}, \facility{Spitzer}.

\bibliographystyle{jphysicsB}
\bibliography{article}

\end{document}